\documentclass[12pt]{iopart}
\usepackage{iopams}

\usepackage{amssymb}
\usepackage{amsfonts}
\usepackage{bbold}
\usepackage{graphicx}
\usepackage{epsfig}
\usepackage{color}
\usepackage{url}
\usepackage{times}
\usepackage{bm}
\usepackage{mathrsfs}
\usepackage[utf8]{inputenc}
\usepackage{hyperref}
\usepackage{enumerate}
\usepackage{amsthm}
\usepackage{verbatim}
\usepackage{cite}
\usepackage{bbm}
\usepackage{stmaryrd}
\usepackage{slashed}
\usepackage{tikz}
\usepackage{caption}
\usepackage{subcaption}
\usepackage{revsymb} 
\usepackage{upgreek}

\def\bmg{{\bm g}}

\def\bme{{\bm e}}

\def\bme{{\bm e}}

\def\bmf{{\bm f}}

\def\bmQ{{\bm Q}}

\def\dotnabla{{\mathring{\nabla}}}
\def\dotg{{\mathring{g}}}

\def\dots{{\mathring{s}}}
\def\dotXi{{\mathring{\Xi}}}
\def\dotd{{\mathring{d}}}
\def\dotPhi{{\mathring{\Phi}}}

\def\dotR{{\mathring{R}}}
\def\s{\mathfrak{s}}

\theoremstyle{plain}
\newtheorem{proposition}{Proposition}

\newtheorem{remark}{Remark}

\newcommand{\beq}{\begin{equation}}
\newcommand{\eeq}{\end{equation}}
\newcommand{\bea}{\begin{eqnarray}}
\newcommand{\eea}{\end{eqnarray}}
\newcommand{\bit}{\begin{itemize}}
\newcommand{\eit}{\end{itemize}}
\newcommand{\ben}{\begin{enumerate}}
\newcommand{\een}{\end{enumerate}}
\newcommand{\nn}{\nonumber}
\newcommand{\dfrac}[2]{{\displaystyle\frac{#1}{#2}}}
\newcommand{\eqref}[1]{(\ref{#1})}

\newcommand{\eqblock}[2]{(\ref{#1}--\ref{#2})}

\newcounter{mnotecount}

\newcommand{\mnotex}[1]
{\protect{\stepcounter{mnotecount}}$^{\mbox{\footnotesize $\bullet$\themnotecount}}$ 
\marginpar{
\raggedright\tiny\em
$\!\!\!\!\!\!\,\bullet$\themnotecount: #1} }

\newcommand{\rh}{r_{+}}
\newcommand{\rc}{r_{-}}

\def\scri{\mathscr{I}}

\begin{document}

\title[The linearised conformal Einstein field equations around a Petrov-type~D spacetime]%
  {The linearised conformal Einstein field equations around a Petrov-type~D spacetime: the conformal Teukolsky equation}

\author{Edgar Gasper\'in}
\address{Instituto de Ciencias Nucleares, Universidad Nacional Aut\'onoma 
de M\'exico,\\ Circuito Exterior s/n, Cd. Universitaria, Cd.~Mx., 04510, M\'exico}

\author{Rodrigo Panosso Macedo}
\address{Center of Gravity, Niels Bohr Institute, Blegdamsvej 17, 2100 Copenhagen, Denmark}

\author{Justin Feng}
\address{CEICO, FZU - Institute of Physics of the Czech Academy of Sciences, Na Slovance 1999/2, 182 21 Prague 8, Czech Republic}

\begin{abstract}
While the Teukolsky equation plays a central role in traditional treatments of perturbations of algebraically special spacetimes, its relation to Friedrich's conformal Einstein field equations (CEFEs) remains largely unexplored. Here we develop a conformal formulation of black-hole perturbation theory based on the CEFEs and derive the conformal Teukolsky equation. Starting from a transparent review of Friedrich's regularisation strategy, this work establishes a direct connection between mainstream curvature-based linear perturbation theory and conformal formulations of general relativity. This perspective is timely given the growing relevance of hyperboloidal frameworks in black-hole perturbation theory, where conformal compactification is introduced at the level of an already linearised effective wave equation. Here instead, the conformal factor is a dynamical variable within the field equations. In the non-linear equations there is a coupling between conformal and curvature perturbations; however, when linearised around a Petrov-type D background, the conformal factor decouples from the equations governing the Newman-Penrose components $\phi_0$ and $\phi_4$ of the rescaled Weyl tensor. The resulting equation preserves the structural form of the classical Teukolsky equation while remaining regular at the conformal boundary. This provides a geometric interpretation of the hyperboloidal master variable and an entry point into the CEFE framework. We further derive the conformal Teukolsky equation for a conformal representation of Kerr spacetime where spatial infinity is realised as a blown-up cylinder. By bridging conformal and traditional approaches to black-hole perturbation theory, the framework highlights a geometrically regular representation of perturbative dynamics that may inform extensions beyond the linear regime.
\end{abstract}

\maketitle

\setlength{\parindent}{0pt}

\section{Introduction}

Given the non-linear nature of Einstein’s field equations (EFEs), studies within the linear regime have long served as a crucial strategy for understanding General Relativity—both in terms of its formal mathematical structure and its astrophysical implications. In this context, gravitational waves (GWs) are among the most significant predictions of the EFEs, confirming the causal structure inherent in the gravitational interaction. Textbooks typically introduce GWs by linearising the EFEs around a flat background, but modern gravitational wave astronomy relies strongly on black hole perturbation theory (BHPT)~\cite{Mag07a}—that is, linearisation schemes around black hole (BH) spacetimes. Remarkably, in a wide range of scenarios, BHPT reduces the EFEs to a single master equation, with the associated master function capturing essential mathematical and physical features necessary for a correct understanding of the problem~\cite{Cha83a}.

\smallskip

The Regge–Wheeler–Zerilli (RWZ) formalism is a well-established approach for deriving master equations governing the radiative degrees of freedom propagating on a spherically symmetric spacetime. This approach considers perturbations of the metric tensor around a spherically symmetric background and, after a judicious gauge choice, reduces the linearised EFEs to decoupled equations for the axial and polar (or equivalently, odd and even) sectors of the metric perturbations \cite{Regge:1957td, Zerilli:1970se,Cha83a}. Despite modern geometrical treatments of the Regge–Wheeler–Zerilli formalism~\cite{Martel:2005ir,Mukkamala:2024dxf}, a \emph{metric formulation} of BH perturbation theory quickly becomes cumbersome and impractical when applied to non-spherically symmetric black hole spacetimes—such as perturbations around the more physically relevant Kerr geometry~\cite{Franchini:2023xhd}.

\smallskip

On the other hand, \emph{curvature formulations} offer a powerful strategy to overcome the limitations of metric-based approaches. Building on the Newman–Penrose formalism~\cite{Newman:1961qr}, Bardeen and Press~\cite{Bardeen:1973xb} considered perturbations of the curvature tensor (as opposed to the metric) around the Schwarzschild spacetime and derived master wave equations for the curvature radiative degrees of freedom, represented by the Weyl scalars $\Psi_0$ and $\Psi_4$ of the Newman-Penrose formalism. This formalism was soon after generalised to the Kerr spacetime, leading to one of the most significant results in BHPT: the Teukolsky equation (TE)~\cite{Teu73}.

\smallskip

The success of BHPT in modeling gravitational radiation from perturbed black holes has made it a cornerstone of GW astronomy, particularly in the era inaugurated by the LIGO–Virgo–KAGRA detections~\cite{LIGO2016,KAGRA:2021vkt,LIGOScientific:2024elc}. However, as the detectors' sensitive increases, so too does the need for more accurate waveform modeling~\cite{Purrer:2019jcp,LISAConsortiumWaveformWorkingGroup:2023arg}, prompting major efforts to extend BHPT beyond the linear regime~\cite{Campanelli:1998jv,Loutrel:2020wbw,Spiers:2023cip}. In parallel, the prospect of testing alternative theories of gravity with upcoming GW observatories has motivated generalisations of the Teukolsky formalism to scenarios beyond the standard EFEs~\cite{Li:2022pcy,Hussain:2022ins,Cano:2023tmv}.

\medskip

Aligned with ongoing efforts to meet the demands of high-precision GW astronomy, the so-called hyperboloidal framework has gained increasing prominence over the past decades~\cite{Zenginoglu:2011jz,Mac20,PanossoMacedo:2023qzp,PanossoMacedo:2024nkw}. It offers a geometric approach for constructing coordinate systems that are optimally suited to describing radiative systems. As a result, it overcomes severe limitations inherent in the standard formulation of BHPT --- such as ill-defined representations of retarded fields, diverging in the asymptotic region --- thereby significantly reducing systematic errors in numerical computations. Additionally, the framework enables the import of powerful tools from the theory of non-self-adjoint operators into gravitational physics~\cite{JarMacShe21,Jaramillo:2022kuv,PanossoMacedo:2025xnf}.

\smallskip

Formally, the hyperboloidal framework builds upon Penrose’s seminal work on the treatment of spacetime infinities in General Relativity~\cite{Pen64}. Using conformal transformations, spacetime infinities are mapped onto well-defined, finite hypersurfaces. Loosely speaking, one works in a conformal spacetime ${\cal M}$ with metric $g_{ab}$, which relates to the physical spacetime $\left( \tilde{\cal M}, \tilde{g}_{ab}\right)$ via a conformal factor $\Xi$ through the relation $\tilde{g}_{ab} = \Xi^{-2} g_{ab}$. The hypersurface defined by $\Xi = 0$ corresponds, formally, to the asymptotic boundary of spacetime~\cite{Val16}. In the context of the hyperboloidal framework, a suitable choice of coordinates is introduced within this compactified setting such that the constant-time hypersurfaces are carefully adapted to the black hole horizon ${\cal H}^+$  and the asymptotic wave zone. i.e. future null infinity $\mathscr{I}^+$.

\medskip

In practice, early implementations of hyperboloidal methods in black-hole perturbation theory often deliberately avoided fully covariant conformal formulations, favouring instead more direct coordinate-based constructions aimed at maintaining accessibility and numerical efficiency (see e.g.~\cite{Zenginoglu:2008uc, Zenginoglu:2011jz}). As a result, the conformal structure typically enters at the level of an already linearised effective wave equation rather than through a conformally regular formulation of the field equations themselves. In particular, the hyperboloidal strategy relies on the existence of an underlying \emph{effective} linear wave-like equation, such as those arising in the Regge–Wheeler–Zerilli or Bardeen–Press–Teukolsky formalisms. Once the coordinates are adapted to hyperboloidal foliations extending to ${\cal H}^+$ and $\mathscr{I}^+$, a formally regular effective equation can be obtained, albeit in a coordinate-dependent manner. Working primarily within the linear regime allows one to prescribe the conformal structure explicitly; for instance, the conformal factor $\Xi$ is often chosen to coincide with a compactified radial coordinate, effectively fixing the conformal compactification through the coordinate construction itself~\cite{Mac20,PanossoMacedo:2023qzp}.

\medskip

This approach, however, stands in contrast to a treatment of the full, non-linear Einstein field equations from first principles. In particular, the non-linear evolution of the EFEs on hyperboloidal slices has proven extremely challenging, as the equations contain formally singular terms at $\mathscr{I}$. Nevertheless, significant progress has been made in recent years toward implementing non-linear hyperboloidal evolutions~\cite{Zen08,BarSarBuc11,Van15,HilHarBug16,VanHus17,Van23,PetGauVanHil24}. Currently, matching Cauchy slices with compactified null slices to include $\scri^+$ on the numerical grid offers the most efficient approach to solve the non-linear Einstein field equations in scenarios relevant to gravitational wave astronomy~\cite{Winicour:2008vpn,Ma:2023qjn}.

\smallskip

One way to understand the origin of the singular terms at $\mathscr{I}$ is to consider a reformulation of the EFEs for the triplet $\left( \mathcal{M}, {g}_{ab}, \Xi \right)$. Upon performing the conformal rescaling $\tilde{g}_{ab} = \Xi^{-2} g_{ab}$, the vacuum Einstein field equations $\tilde{G}_{ab}[\tilde{g}] = 0$ transform into $G_{ab}[g] = T_{ab}[\Xi]$, where $\tilde{G}_{ab}$ and $G_{ab}$ denote the Einstein tensors associated with the physical and conformal metrics, respectively, and $T_{ab}[\Xi]$ is an effective “energy–momentum” tensor constructed from the conformal factor $\Xi$~\cite{Zen08}. However, $T_{ab}[\Xi] \sim \Xi^{-2}$ near $\Xi = 0$, which imposes stringent regularity conditions for obtaining smooth solutions at the asymptotic boundary of the spacetime.
This makes the numerical evolution of the full non-linear (singular) hyperboloidal formulations particularly challenging, as evidenced by the fact that, to date, only spherically symmetric evolutions have been achieved ---see \cite{Van23,PetGauVanHil24} for the most recent developments.

\smallskip
 
Nevertheless, the standard (off-the-shelf) theory of partial differential equations (PDE) does not apply to these singular equations. This leads to the pursuit of \emph{formally regular tensorial equations} for the triplet $\left( \mathcal{M}, g_{ab}, \Xi \right)$, for which, once a coordinate system is introduced, standard PDE theory—e.g., as in~\cite{Kat75}—can be directly applied. Crucially, this is possible if both the conformal factor $\Xi$ and curvature quantities are treated as dynamical variables. The resulting system of equations is known as the Conformal Einstein Field Equations (CEFEs), originally derived by H. Friedrich in \cite{Fri81a}. The CEFE formulation of the EFEs is particularly well-suited for studying the global structure of spacetimes, with applications ranging from rigorous mathematical results on spacetime stability~\cite{Fri86b, Fri86c, Fri95, LueVal09, GasVal17, HilValZha20} to fully non-linear, three-dimensional numerical simulations~\cite{FraSteThw25, FraSte21, FraSte23, FraSte24}.

\smallskip

It should be stressed that, although the CEFEs are formally regular at null infinity ($\mathscr{I}$), where $\Xi = 0$ but $\nabla_a \Xi \neq 0$, the initial data for the CEFEs becomes singular at spatial infinity ($i^0$), where both $\Xi = 0$ and $\nabla_a \Xi = 0$. This region lies outside the scope of hyperboloidal foliations and presents challenges that are conceptually distinct from those encountered at null infinity~\cite{Ger72,Fri98a,HinVas17}.

\smallskip

As originally emphasised by Penrose, the conformal structure of asymptotically flat spacetimes with non-zero mass becomes degenerate at spatial infinity $i^0$ \cite{Pen65}. Within conformal formulations of general relativity, this difficulty is often addressed through “blow-up’’ constructions in which the point $i^0$ is replaced by an extended geometric structure, allowing a regular treatment of the asymptotic region. One prominent example is Friedrich’s cylinder at spatial infinity \cite{Fri98a}, constructed using conformal geodesics and adapted coordinate systems. Although explicit constructions of such representations remain technically challenging for rotating spacetimes such as Kerr, Ref.~\cite{Hennig:2020rns} introduced an alternative cylinder-like representation of $i^0$ based on coordinates adapted to null geodesics, which enables the study of massless scalar propagation near spatial infinity and will serve as one of the asymptotic settings considered in this work.

\medskip

In view of the challenges associated with treating spacetime infinities in a mathematically rigorous and numerically stable way for the full non-linear Einstein field equations --- especially when compared with the  simpler, coordinate-dependent treatment in the linearised setting --- this work develops a connection between mainstream approaches in BHPT and the CEFEs. Specifically, we derive the conformal Teukolsky equation from first principles within the CEFE framework. The motivation for pursuing a CEFE-based Teukolsky equation is twofold:

\smallskip

First, from the perspective of developing a linear black hole perturbation theory rooted in the CEFEs, it remains unclear whether the operations of linearisation and conformal compactification commute. This issue is highlighted by the absence versus presence of formally singular terms in the linear versus non-linear formulations of the hyperboloidal framework. 

\smallskip
Second, from a conceptual standpoint, the conformal factor in the CEFE formalism is a dynamical quantity --- unlike in the linear hyperboloidal approach, where the conformal structure of spacetime is fixed a priori. These motivations are particularly relevant in light of recent efforts to extend BHPT beyond the linear regime, aiming for regular and accurate results both at the black hole horizon and in the asymptotic wave zone, key requirements for precision gravitational-wave astronomy. 

\smallskip

Beyond the primary hyperboloidal setting, we also demonstrate that the CEFE-Teukolsky formalism extends to the more challenging asymptotic regime near spatial infinity $i^0$, deriving the equation using a blown-up cylinder representation introduced in~\cite{Hennig:2020rns}.

\smallskip

The remainder of this paper is structured as follows. In Section~\ref{sec:CEFE_Review}, we review the CEFE formalism, summarising its core equations and recent results in the linear regime. Section~\ref{sec:CEFE-Teukolsky} then presents a derivation of the conformal Teukolsky equation directly from the CEFEs, highlighting the roles played by the so-called rescaled Weyl tensor and the Killing spinor equation. Finally, in Section~\ref{sec:Results}, we present our results. We not only reproduce expressions for the Teukolsky equation previously derived using linear hyperboloidal foliations, but we also formulate, for the first time, the conformal Teukolsky equation in a cylinder-like representation of $i^0$. Section~\ref{sec:Conclusion} concludes the paper with a discussion of implications and future directions.

\subsection{Notation}
Given the central role played by the conformal manifold in our approach, we adopt notational conventions that are standard in the literature on the CEFEs. The physical spacetime and all tensorial quantities defined on it are denoted with a tilde; for instance, $(\tilde{\mathcal{M}}, \boldsymbol{\tilde{g}})$ denotes the physical spacetime, and $\tilde{Q}$ represents a physical observable. The unphysical\footnote{The term “unphysical” to describe the conformally rescaled spacetime originates from~\cite{Pen64}. We retain this terminology not only for historical consistency but also to clearly distinguish quantities directly associated with the conformal metric from others that could undergo additional conformal rescaling.} spacetime and the corresponding conformally rescaled quantities are denoted by $(\mathcal{M}, \boldsymbol{g}, \Xi)$ and $Q$, respectively. We further use $\mathring{Q}$ to refer to background quantities and $\check{Q}$ to denote first-order perturbations. A summary of this notation is provided in Table~\ref{Tab:Notation}.

Lowercase Latin letters are used for abstract tensor indices, while Greek letters refer to components in a specific coordinate system. For instance, $\ell^a$ and $\hat{\ell}^a$ represent different vector fields, whereas $\ell^\mu$ and $\ell^{\hat{\mu}}$ are the components of the same abstract vector $\ell^a$ in the coordinate systems $x^\mu$ and $\hat{x}^\mu$, respectively. 
Symmetrisation and antisymmetrisation of indices is denoted with round and square brackets, for example $T_{(ab)}: =(T_{ab} +T_{ba})/2, \;\; T_{[ab]}:=(T_{ab} -T_{ba})/2$. Similarly, the trace-free part is denoted with curly brackets,  for example $T_{\{ ab \} } := T_{ab} - \dfrac{1}{4}T_{c}{}^c g_{ab} $. Complex conjugation is indicated by $Q^\ast$.
Throughout this work, we adopt geometrised units in which $c = G = 1$.

\begin{table*}[h!]
    \centering
        \caption{Notation and conventions summary}
    \begin{tabular*}{0.90\textwidth}{|c|l |c| c |l|}
        \cline{1-2} \cline{4-5}
        \textbf{Spacetime} & \textbf{Description} & & \textbf{Qty.} & \textbf{Description} \\[2pt]
        \cline{1-2} \cline{4-5}
        \multicolumn{5}{c}{} \\[-8pt]
        \cline{1-2} \cline{4-5}
        $~$ & $~$ & & $~$ & $~$ \\[-8pt]
        $(\tilde{\mathcal{M}},\tilde{\bf g})$ & Physical spacetime & & $\tilde{Q}$ & Physical $Q$ \\[8pt]
        $(\mathcal{M},\bf g, \Xi)$ & Unphysical spacetime & & ${Q}$ & Unphysical $Q$ \\[4pt]
        $(\mathring{\tilde{\mathcal{M}}},\mathring{\tilde{\bf g}})$ & Phys. background spacetime  & & $\mathring{Q}$ & Background value for $Q$ \\[4pt]
        $(\mathring{\mathcal{M}},\mathring{\bf g}, \mathring{\Xi})$ & Unphys. background spacetime & & $\check{Q}$ & Perturbation of $Q$ \\[4pt]
        \cline{1-2} \cline{4-5}
    \end{tabular*}
    \label{Tab:Notation}
\end{table*}
\section{The Conformal Einstein Field Equations} \label{sec:CEFE_Review}
\subsection{The CEFEs as a system of tensorial equations} 
The conformal formulation of General Relativity begins by considering two manifolds, $\mathcal{M}$ and $\tilde{\mathcal{M}}$, equipped with metrics $\bmg$ and $\tilde{\bmg}$, respectively, which are conformally related:
\begin{eqnarray}
\label{eq:ConformalTransformation}
g_{ab} := \Xi^2 \tilde{g}_{ab}.
\end{eqnarray}
The pair $(\tilde{\mathcal{M}}, \tilde{\bmg})$ is assumed to satisfy the Einstein field equations (EFEs) and is commonly referred to as the physical spacetime in the literature. In contrast, the conformally rescaled pair $(\mathcal{M}, \bmg)$ is known as the unphysical spacetime, a term that reflects its role as a formal mathematical construction. The conformal factor then determines the boundary $\partial \mathcal{M}$ of the unphysical spacetime as the region where $\Xi = 0$.
To find the governing equations for $(\mathcal{M}, \bmg)$ one starts by observing that if $\tilde{\bmg}$ solves the (vaccum) EFEs $\tilde G_{ab}[\tilde{\bmg}] = 0$ then the unphysical metric $\bmg$ satisfies:
\begin{eqnarray}
\label{eq:formallysingular}
G_{ab}[{\bmg}] = -\dfrac{2}{\Xi} \left( \nabla_a \nabla_b \Xi - g_{ab} \nabla^c\nabla_c \Xi\right) - \dfrac{3}{\Xi^2}g_{ab} \nabla_c \Xi \nabla^c \Xi.
\end{eqnarray}
---see e.g. Ref.~\cite{Zen08}. Notice that the equations pick up explicitly singular terms $\Xi^{-1}$ and $\Xi^{-2}$. 

The CEFEs arises as a reformulation of the EFEs with the property that they are formally regular in the conformal factor $\Xi$ ---i.e. without $\Xi^{-|p|}$  terms. This is of interest for studying the global structure of solutions to the EFEs as it allows to compute quantities
directly at the conformal boundary $\Xi=0$.

In order to be self-contained and benefit of the reader not familiar with the CEFEs, here we summarise  the main steps in the derivation of the CEFEs. The key idea for regularising equation \eqref{eq:formallysingular} is not to read it as an equation for $\bmg$ but rather as an equation for $\Xi$, hence effectively promoting $\Xi$ to be a dynamical quantity.
To obtain the equation for $\Xi$
one starts by observing that,
the Schouten tensor $L_{ab}$, defined as
\begin{equation}
L_{ab}:=\dfrac{1}{2}R_{ab}+\dfrac{1}{12}R g_{ab}, 
\end{equation}
under conformal rescalings of the metric
\eqref{eq:ConformalTransformation}, transforms as follows: 
\begin{eqnarray}\label{eq:ConformalTransformationSchouten}
    L_{ab}-\tilde{L}_{ab} = -\Xi^{-1}\nabla_a\nabla_b\Xi+ \frac{1}{2}g_{ab}\Xi^{-2}\nabla_c\Xi\nabla^c\Xi.
\end{eqnarray}
If the vacuum Einstein field equations hold $\tilde{G}_{ab}=0$, equivalently $\tilde{R}_{ab}=0$, then one has $\tilde{L}_{ab}=0$.
Substituting $\tilde{L}_{ab}=0$  in 
equation \eqref{eq:ConformalTransformationSchouten}
and solving for $\nabla_a\nabla_b \Xi$
one obtains
 \begin{eqnarray}\label{eq:CFE_tensor_zeroquants_1}
   && \nabla_{a}\nabla_{b}\Xi +\Xi L_{ab} - s g_{ab}=0,  
  \end{eqnarray}
where $s$ is a scalar defined as
\begin{equation}
 s:=\dfrac{1}{4}\nabla_c\nabla^c\Xi+\dfrac{1}{24}R\Xi.
 \end{equation}
 The field $s$ in known as the Friedrich scalar.
 Notice that equation \eqref{eq:CFE_tensor_zeroquants_1}
 is a manifestly regular version (there are no $\Xi^{-1}$ factors) of the formally singular equation \eqref{eq:formallysingular}. The price of this regularisation is that now $\Xi$ and $s$ and $L_{ab}$ are to be thought of as independent dynamical quantities. The equation for $\Xi$ is of course \eqref{eq:CFE_tensor_zeroquants_1} and now, one needs to find equations for $s$ and $L_{ab}$.
  The equation for $s$ comes as an integrability condition for equation \eqref{eq:CFE_tensor_zeroquants_1}:
 taking a derivative of \eqref{eq:CFE_tensor_zeroquants_1}
 and commuting covariant derivatives one arrives at
 \begin{eqnarray}
    && \nabla_{a}s +L_{ac} \nabla ^{c}\Xi=0. \label{standardCEFEs}
  \end{eqnarray}

In order to find an equation for $L_{ab}$ one looks at the second Bianchi identities, which in terms of the Schouten and Weyl tensors read:
\begin{eqnarray}\label{eq:BianchiL}
    \nabla_{c}L_{db}-\nabla_{d}L_{cb}= \nabla_{a}C^{a}{}_{bcd},
\end{eqnarray}
where $C^{a}{}_{bcd}$ is the Weyl tensor.
If on the one hand we have found an equation for $L_{ab}$, on the other hand we have introduced $C{}^a{}_{bcd}$ as another variable. Hence, one needs an equation for $C^{a}{}_{bcd}$ to close the system. 

Using the conformal transformation formulae for the Weyl tensor and its derivatives, one obtains the identity
\begin{eqnarray}\label{eq:WeylBianchi}
    \nabla_{a}(\Xi^{-1}C^{a}{}_{bcd})=\Xi^{-1}\tilde{\nabla}_a 
    \tilde{C}^{a}{}_{bcd}.
\end{eqnarray}
The right-hand-side of the above equation vanishes, thanks to the physical version of the second Bianchi identity \eqref{eq:BianchiL}, i.e. 
$\tilde{L}_{ab}=0\Longrightarrow\tilde{\nabla}_a 
    \tilde{C}^{a}{}_{bcd}=0$.
The left-hand side of equation \eqref{eq:WeylBianchi} then suggests to introduce the so-called \emph{rescaled Weyl tensor}, defined as
    \begin{eqnarray}
 \label{eq:d}
d^{a}{}_{bcd}:=\Xi^{-1}C^{a}{}_{bcd},
  \end{eqnarray}
  so that equations \eqref{eq:WeylBianchi} and \eqref{eq:BianchiL} read:
  \begin{eqnarray}
 &&  \nabla_{a}L_{bc}-\nabla_{b}L_{ac} -  d^{d}{}_{cab}\nabla_d{}\Xi=0 ,  \label{standardCEFESchouten}\\ 
 &&   \nabla_{e}d^{e}{}_{abc}=0.
    \end{eqnarray}

The rescaled Weyl tensor is an object that will play a central role in the discussion of this paper. Therefore, some comments are in order. 

The Weyl tensor satisfies $C^{a}{}_{bcd}=\tilde{C}^{a}{}_{bcd}$,
stressing that the conformal invariance  property is only valid for this particular arrange of covariant/contravariant indices. With other index positions one needs to take into account that $g_{ab}=\Xi^2\tilde{g}_{ab}$ and recall that indices of tilded quantities are moved only with $\tilde{g}_{ab}$. Indeed, a representation in terms of only covariant indices yields the following relation between the rescaled, unphysical and physical Weyl tensors:
\begin{eqnarray}
\label{unphysical_Weyl}
	d_{abcd}:=\Xi^{-1}C_{abcd} = \Xi \tilde{C}_{abcd}.
\end{eqnarray}

While the definition of the rescaled Weyl tensor  arises directly from the mathematical identity \eqref{eq:WeylBianchi}, it also reflects a clear physical idea: the radiation field should remain finite at infinity ($\Xi = 0$). Indeed, if the physical Weyl tensor $\tilde{C}^{a}{}_{bcd}$ vanishes sufficiently fast near $\Xi = 0$, then dividing by $\Xi$ leads one to expect
$d^{a}{}_{bcd} = \Xi^{-1} \tilde{C}^{a}{}_{bcd} \simeq \mathcal{O}(1)$
as the conformal boundary is approached. This re-scalling is utterly akin to the well-known asymptopic behaviour of a scalar field $\tilde\Psi \sim 1/r$, which leads to the conformal re-scalling $\Psi=\Xi^{-1}\tilde \Psi$. 

\medskip

Overall, with this procedure one obtains the following set of tensorial equations encoding in a formally regular way (no $\Xi^{-1}$ terms) the formally singular equation  \eqref{eq:formallysingular}:
\label{CFE_tensor_zeroquants}
\begin{eqnarray}
\label{CFE_tensor_zeroquants_XiEq}
 && \nabla_{a}\nabla_{b}\Xi +\Xi L_{ab} - s g_{ab} =0 ,\label{StandardCEFEsecondderivativeCF}\\ &&
\nabla_{a}s +L_{ac} \nabla ^{c}\Xi=0 , \label{standardCEFEs}\\ &&
\nabla_{a}L_{bc}-\nabla_{b}L_{ac} - d^{d}{}_{cab}\nabla_d{}\Xi =0 ,
 \label{standardCEFESchouten}\\ && 
 \nabla_{e}d^{e}{}_{abc} =0.\label{standardCEFERescWeyl}
\end{eqnarray}

\subsection{The CEFEs as a system of partial differential equations} 
Given the system  \eqref{CFE_tensor_zeroquants_XiEq}-\eqref{standardCEFERescWeyl}
a natural question that arises is, what are then the equations for the metric?
To clarify this point,  it should be emphasised that equations \eqref{CFE_tensor_zeroquants_XiEq}-\eqref{standardCEFERescWeyl} are tensor relations for the conformal fields
$\{\Xi, s, L_{ab}, d^{a}{}_{bcd}\}$. In other words, they are at the same level of abstraction as the physical vacuum EFEs when expressed as $\tilde{R}_{ab}=0$.  At this level, the physical EFEs $\tilde{R}_{ab}=0$ is an algebraic equation for the physical Ricci tensor while the CEFEs are, at its core, differential conditions satisfied by the conformal (unphysical) curvature $\{ L_{ab}, d^{a}{}_{bcd}\}$. The process of extracting a PDE-system (evolution and constraints) out of an abstract tensorial equation is known as \emph{ hyperbolic reduction} ---because the targeted PDE-type for the evolution equations to be extracted is hyperbolic.

\smallskip
In the case of the (physical) EFES  $\tilde{G}_{ab}=0$ there are different hyperbolic reduction strategies. For instance, in metric hyperbolic reductions,  
$\tilde{G}_{ab}=0$ is read as a system of equations for the metric components $\tilde{g}_{\mu\nu}$ which, if coordinates $\tilde{x}^\mu$ are chosen appropriately, are of hyperbolic type.
\emph{Metric hyperbolic reductions} are particularly popular in Numerical Relativity, such as the \emph{GHG, BSSN, Z4c} formulations to mention a few.

\smallskip
Alternatively, \emph{non-metric hyperbolic reductions} take as fundamental variable a tetrad and the connection coefficients, which are linked to the curvature ---and hence $\tilde{R}_{ab}=0$--- through the Cartan structure equations and Bianchi identities.
Examples in this category are the \emph{Newman-Penrose}  or \emph{Christodolou-Klainerman} formulations of the (physical) EFEs ---see \cite{Newman:1961qr, ChrKla93}. 

\smallskip
Similarly, and analgously with the physical case, for the CEFEs there are several different hyperbolic reductions available, which augment the system \eqref{StandardCEFEsecondderivativeCF}-\eqref{standardCEFERescWeyl} with equations for either the unphysical metric $g_{\mu\nu}$ or a tetrad and connection-coefficients.

\smallskip
In the case of metric hyperbolic reductions of the CEFEs, the first step is to obtain a system of wave equations for the field variables $\{ \Xi, s, L_{ab}, d^a{}_{bcde}\}$. By taking the trace of equation~\eqref{CFE_tensor_zeroquants_XiEq}, hitting the remaining equations \eqref{standardCEFEs}-\eqref{standardCEFERescWeyl}  with the derivative operator, a direct manipulation yields
\label{geo_wave}
\begin{eqnarray}
  \square \Xi  &=& 4 s - \dfrac{1}{6} \Xi R \label{geowave1},\\
  \square s &=& \Xi  \Phi _{ab} \Phi ^{ab}
  - \dfrac{1}{6} \nabla _{a}R \nabla ^{a}\Xi
  - \dfrac{1}{6} s R + \big(\dfrac{R}{12}\big)^2 \Xi
  \\ \square \Phi_{ab} &=& 4 \Phi _{a}{}^{c} \Phi _{bc}
  - \Phi _{cd} \Phi ^{cd} g_{ab}
  -2 \Xi  \Phi ^{cd} d_{acbd} + \nn \\
  &&\dfrac{1}{3} \Phi _{ab} R
  + \dfrac{1}{6} \nabla _{b}\nabla _{a}R
  - \dfrac{1}{24} g_{ab} \square R
  \\ \square d_{abcd} &=&  \dfrac{1}{2} d_{abcd} R
  + 2 \Xi  d_{a}{}^{e}{}_{d}{}^{p} d_{becp}
  -2 \Xi  d_{a}{}^{e}{}_{c}{}^{p} d_{bedp}
  -2 \Xi  d_{ab}{}^{ep} d_{cedp} \label{geowave4}
\end{eqnarray}
where the Schouten tensor has been exchanged for the trace-free part of the Ricci tensor
  \begin{eqnarray}
    \Phi_{ab} &:=& \frac{1}{2}\Big(R_{ab} - \frac{1}{4}Rg_{ab}\Big) 
            =   L_{ab} -  \frac{5}{24}Rg_{ab}.	
  \end{eqnarray}
Then, the equation for the metric schematically reads
\begin{equation}
    (\partial^2 g)_{\mu\nu} + (\partial g)^2_{\mu\nu} = L_{\mu\nu} \label{eq:metric_eq_schematic}
\end{equation}
where $ (\partial^2 g)_{\mu\nu}$ and $(\partial g)_{\mu\nu}$ denote second order and first order derivatives of the metric components, respectively, with the Schouten tensor $L_{\mu\nu}$ an independent source governed by the system \eqref{CFE_tensor_zeroquants_XiEq}-\eqref{standardCEFERescWeyl}, or equivalently \eqref{geowave1}-\eqref{geowave4}.

To ensure that equation \eqref{eq:metric_eq_schematic} is hyperbolic one has to choose coordinates $x^\mu$ appropriately ---effectively mimicking the traditional metric hyperbolic reductions of the physical EFEs with a source. Formally, the equation for the unphysical metric looks formally identical to the evolution equations in the Generalised Harmonic Gauge formation of the standard physical Einstein field equations:
  For further details, the interested reader is referred to \cite{Pae13} or the Appendix of \cite{FenGas23} where these equations are written explicitly in this format.

In the next section, we review the current approach for perturbation theory in the CEFEs, which is formulated in the context of metric hyperbolic reductions. Then, in Sec.~\ref{sec:CEFE-Teukolsky} we turn our attention to non-metric hyperbolic reductions of the CEFEs, which turn out to be more appropriate to derive the conformal Teukolsky equation.

\subsection{Linearised Conformal Einstein Field Equations}

The CEFEs have been primarily used for studying global non-linear perturbations, while the study of linear perturbations using the CEFEs has been relatively less exploited. In Ref. \cite{FenGas23} the
metric and second order hyperbolic reduction of the CEFEs described before, was linearised around an arbitrary
 conformal background $(\mathring{\mathcal{M}},\mathring{\bmg}, \mathring{\Xi})$. 

 The linear system of equations takes the schematic form
\label{eq:LinearSystem}
\begin{eqnarray}
    \mathring{\square} \check{g}_{\mu\nu}
    &=
    H^{g}_{(\mu\nu)}(\check{\bm\phi},
    \dotnabla\check{\bm\phi} \; ; \;  \mathring{\bm\phi}, \bmf, 
    \dotnabla\mathring{\bm\phi},
    \dotnabla \bmf ), \label{eq:LinearSystem1}
    \\
    \mathring{\square} \check{\Xi} 
    & = H^{\Xi}_{}(\check{\bm\phi}, \dotnabla \check{\bm\phi}
    \; ; \; 
    \mathring{\bm\phi}, \bmf, \dotnabla\mathring{\bm\phi} ),  \\
    \mathring{\square} \check{s} 
    & = 
    H^{s}(\check{\bm\phi}, \dotnabla\check{\bm\phi}
    \; ; \; 
    \mathring{\bm\phi}, \bmf, \dotnabla\mathring{\bm\phi},
    \dotnabla \bmf ) ,  \\ 
    \mathring{\square} \check{\Phi}_{\mu\nu }
    &= 
    H^{\Phi}_{(\mu\nu)}(\check{ \bm\phi},
    \dotnabla \check{ \bm\phi} \; ; \; 
    \mathring{\bm\phi}, \bmf, \dotnabla\mathring{\bm\phi},
      \dotnabla \bmf,  \dotnabla\dotnabla \bmf ),
    \\
    \label{eq:Lin_d}
    \mathring{\square} \check{ d}_{\mu\nu\alpha\beta}
    & =
    H^{W}_{[\mu\nu][\alpha\beta]}(\check{ \bm\phi}, 
    \dotnabla \check{ \bm\phi}
    \; ; \; 
    \mathring{\bm\phi}, \bmf, \dotnabla\mathring{\bm\phi},
    \dotnabla\dotnabla\mathring{\bm\phi}, \dotnabla \bmf ).
  \end{eqnarray}
  As summarised in Table~\ref{Tab:Notation}, $\mathring{\bmQ}$ and $\check{\bmQ}$ denotes the respective background value and perturbation of a particular quantity $\boldsymbol{Q}$. For instance, $\dotnabla$ denotes the Levi-Civita covariant derivative respect to the conformal background metric $\dotg_{\mu\nu}$. The left hand side contains the wave operators $\mathring{\square}: = \mathring{g}^{\mu\nu}\mathring{\nabla}_\mu\mathring{\nabla}_\nu$ applied on the perturbed dynamical variables $\check{ \bm\phi} = \{      \check{g}_{\mu\nu},      \check{\Xi} ,  \check{s}, \check{\Phi}_{\mu\nu },\check{ d}_{\mu\nu\alpha\beta}  \}$, whereas the right-hand side combines lower order derivatives of the dynamical variables $\check{ \bm\phi}$, gauge source functions $\bmf$, as well as known quantities derived directly from the background solution $\mathring{ \bm\phi} = \{      \mathring{g}_{\mu\nu},      \mathring{\Xi} ,  \mathring{s}, \mathring{\Phi}_{\mu\nu },\mathring{ d}_{\mu\nu\alpha\beta}  \}$.  
  The linear system of equations \eqref{eq:LinearSystem1}-\eqref{eq:Lin_d} were derived in the so-called generalised Lorenz Gauge ($\dotnabla_\beta \check{g}^{\alpha\beta}=F^\alpha$)
 which can be thought as the linear analogue of the generalised harmonic gauge condition ($\Gamma^{\alpha}=-\mathcal{H}^\alpha$). 
 For the interested reader, the functions $H$ on the right-hand side are explicitly given in Appendix A of \cite{FenGas23}.

Equation~\eqref{eq:Lin_d} can be interpreted as a generalised representation of the conformal Teukolsky equation in an arbitrary background. Indeed, it provides an explicit wave equation for components of the (rescaled) Weyl tensor and, as such, should encode the Teukolsky equation as a particular case.
Most importantly, the presence of generic lower order terms $\dotnabla\check{\bm\phi}$ and $\check{\bm\phi}$ in the arguments of the functions $H$ indicates that the linearised CEFEs still provide a coupled system of equations for perturbed dynamical variables $\check{ \bm\phi}$.

After deriving the linearisation scheme around a general background, Ref.~\cite{FenGas23} focused on applications where the background metric is restricted to Minkowski spacetime. In this case, equation~\eqref{eq:Lin_d} decouples because the background rescaled Weyl vanishes, i.e.  $\mathring{d}^{\mu}{}_{\nu\alpha\beta} = 0$ ---see expression for $H^{W}_{\mu\nu\alpha\beta}$ below.  For general backgrounds, however, the background rescaled Weyl tensor does not vanish ($\mathring{d}^{\mu}{}_{\nu\alpha\beta} \neq 0$) and the decoupling does not happen explicitly.
Take for instance equation~\eqref{eq:Lin_d} dictating the dynamics of the perturbed rescaled Weyl tensor $\check{ d}_{\mu\nu\alpha\beta}$. Setting all the Lorenz gauge source functions to zero to simplify the expression as much as possible, its right-hand-side reads
\begin{eqnarray}   
&H^{W}_{\mu  \nu  \alpha  \beta }  = \dfrac{1}{2} \dotR \check{d}_{\mu  \nu  \alpha  \beta  } - 
 (4 \dotPhi ^{\delta  \sigma  } \check{g}_{\delta  \sigma  }  + \dfrac{1}{12} \dotR \check{g}^{\delta  }{}_{\delta  }) \dotd_{\mu  \nu  \alpha  \beta  } - (2 \dotPhi ^{\delta  \sigma  } \check{g}_{\alpha  \delta  }  +6 \dotPhi _{\alpha  }{}^{\delta  } \check{g}_{\delta  }{}^{\sigma  }) \dotd_{\mu  \nu  \beta  \sigma  } \nn \\
&+ (4 \check{\Phi} _{\alpha  }{}^{\delta  }   - \dfrac{1}{6} \dotR \check{g}_{\alpha  }{}^{\delta  } + 2 \dotPhi _{\alpha  }{}^{\delta  } \check{g}^{\sigma  }{}_{\sigma  }) \dotd_{\mu  \nu  \beta  \delta  }   -2 \dotXi  \check{d}_{\alpha  }{}^{\delta  }{}_{\beta  }{}^{\sigma  } \dotd_{\mu  \nu  \delta  \sigma  } \nonumber\\
 &+ \dotd_{\alpha  \delta  \beta  \sigma  } (-2 \dotXi  \check{d}_{\mu  \nu  }{}^{\delta  \sigma  } -2 \check{\Xi}  \dotd_{\mu  \nu  }{}^{\delta  \sigma  })  + (4 \check{\Phi} _{\mu  }{}^{\delta  }  - \dfrac{1}{6} \dotR \check{g}_{\mu  }{}^{\delta  } + 2 \dotPhi _{\mu  }{}^{\delta  } \check{g}^{\sigma  }{}_{\sigma  }) \dotd_{\nu  \delta  \alpha  \beta  } \nonumber\\
& + (-8 \dotXi  \check{d}_{\mu  }{}^{\delta  }{}_{\alpha  }{}^{\sigma  }  -4 \check{\Xi}  \dotd_{\mu  }{}^{\delta  }{}_{\alpha  }{}^{\sigma  }) \dotd_{\nu  \delta  \beta  \sigma  }   + \dotXi  \check{g}^{\delta  \sigma  } 
(2\dotd_{\beta  \delta  \sigma  \lambda} \dotd_{\mu  \nu  \alpha  }{}^{\lambda} + 2 \dotd_{\mu  }{}^{\lambda}{}_{\alpha  \beta  } \dotd_{\nu  \delta  \sigma  \lambda}) \\
&- (6 \dotPhi _{\mu  }{}^{\delta  } \check{g}_{\delta  }{}^{\sigma  } + 2 \dotPhi ^{\delta  \sigma  } \check{g}_{\mu  \delta  }) \dotd_{\nu  \sigma  \alpha  \beta  } \nonumber + 4 \dotXi  \check{g}^{\delta  \sigma  } (\dotd_{\alpha  \sigma  \beta  \lambda} \dotd_{\mu  \nu  \delta  }{}^{\lambda} + \dotd_{\mu  }{}^{\lambda}{}_{\alpha  \delta  } \dotd_{\nu  \lambda\beta  \sigma  } + \dotd_{\mu  \delta  \alpha  }{}^{\lambda} \dotd_{\nu  \sigma  \beta  \lambda}) 
\nonumber\\
& -\dotXi  \check{g}^{\delta  }{}_{\delta  } ( \dotd_{\alpha  \sigma  \beta  \lambda} \dotd_{\mu  \nu  }{}^{\sigma  \lambda}  +2 \dotd_{\mu  }{}^{\sigma  }{}_{\alpha  }{}^{\lambda} \dotd_{\nu  \sigma  \beta  \lambda})  -2 \dotnabla _{\alpha  }\check{g}^{\delta  \sigma  } \dotnabla _{\sigma  }\dotd_{\mu  \nu  \beta  \delta  }  -2 \dotnabla _{\mu  }\check{g}^{\delta  \sigma  } \dotnabla _{\sigma  }\dotd_{\nu  \delta  \alpha  \beta  } \nonumber\\
& + \check{g}^{\delta  \sigma  } \dotnabla _{\sigma  }\dotnabla _{\delta  }\dotd_{\mu  \nu  \alpha  \beta  }   + 2 (\dotnabla _{\delta  }\dotd_{\mu  \nu  \beta  \sigma  }  - \dotnabla _{\sigma  }\dotd_{\mu  \nu  \beta  \delta  }) \dotnabla ^{\sigma  }\check{g}_{\alpha  }{}^{\delta  }  \nonumber\\
&  - ( \dotnabla _{\alpha  }\dotd_{\mu  \nu  \beta  \sigma  } +  \dotnabla _{\mu  }\dotd_{\nu  \sigma  \alpha  \beta  } + 2 \dotnabla _{\sigma  }\dotd_{\mu  \nu  \alpha  \beta  }) \dotnabla ^{\sigma  }\check{g}^{\delta  }{}_{\delta  } + 2 (\dotnabla _{\delta  }\dotd_{\nu  \sigma  \alpha  \beta  }  - \dotnabla _{\sigma  }\dotd_{\nu  \delta  \alpha  \beta  }) \dotnabla^{\sigma  }\check{g}_{\mu  }{}^{\delta  }. \nn
\label{eq:weyltensor}
\end{eqnarray}
Notice, in particular, that the perturbation of the unphysical metric $\check{g}_{\mu\nu}$, the unphysical trace-free Ricci tensor $\check{\Phi}_{\mu\nu}$, and the conformal factor $\check{\Xi}$ appear explicitly in expression~\eqref{eq:weyltensor}.
As a result, even when a black hole background $(\mathring{\tilde{\mathcal{M}}}, \mathring{\tilde{\bmg}})$ and a corresponding conformal extension $(\mathring{\mathcal{M}}, \mathring{\bmg}, \mathring{\Xi})$ are specified, it is not immediately clear whether the equation governing $\check{d}^{\mu}{}_{\nu\alpha\beta}$ decouples from the other variables.

We strongly emphasise that the perturbation of the conformal factor $\check{\Xi}$ enters in these expressions so a priori one would expect a coupling between the perturbations of the rescaled Weyl tensor and the equation for $\check{\Xi}$. In the following sections we investigate the separability of the linearised CEFEs for the rescaled Weyl tensor. In doing so, we derive the conformal Teukolsky equations directly from the CEFE framework.

\section{The CEFE-Teukolsky equation}\label{sec:CEFE-Teukolsky}
Before deriving explicitly the conformal Teuskolsky equations from the linearised system of CEFEs, we recall how the null tetrad behaves under conformal transformations.

\subsection{Conformal transformations of the Newman-Penrose quantities}
Given a set of null tetrad $\{\tilde{\ell}^a, \tilde{k}^a, \tilde{m}^a, \tilde{m}^{*a}\}$ associated with the physical metric $\tilde{g}_{ab}$, the most generic set of unphysical null tetrads $\{ {\ell}^a, {k}^a, {m}^a, {m}^{*a}\}$ aligned with the physical tetrads is given by
\begin{eqnarray}
\label{tetrad_trans_general}
\ell^a=\Xi^{-2-2\omega} b\, \tilde{l}^a, \qquad
k^a=\Xi^{2\omega} b^{-1}\, \tilde{k}^a, \qquad 
m^a = \Xi^{-1} e^{i\vartheta } \tilde{m}^a.
\end{eqnarray}
The above equation accounts not only for the spin/boost transformations encoded by the $(b,\vartheta)$ and inherent to any null tetrad, but they also entail a generic conformal re-scaling compatible with equation~\eqref{eq:ConformalTransformation}. In particular, the parameter $\omega$ incorporates different ways to distribute the conformal factor $\Xi$ within the unphysical tetrad legs $\{\ell^a, k^a \}$ .

\smallskip

\smallskip

Let $\tilde{\Psi}_{n}$ and $\Psi_{n}$ with $n ={0,1,2,3,4}$ denote the independent components (Weyl scalars) of $\tilde{C}_{abcd}$  and $C_{abcd}$ --- cf.~equation~\eqref{unphysical_Weyl} --- with respect 
to the physical and unphysical tetrads accordingly. Additionally, let $\phi_{n}$ denote the analogous components for the rescaled Weyl tensor $d_{abcd}$ --- cf.~equation~\eqref{eq:d} --- with respect to the unphysical tetrad. Then, a direct calculation renders $\phi_n = \Xi \Psi_n$, and 
\begin{eqnarray}
\label{eq:Pelling}
     & \tilde{\Psi}_{0} =\Xi^{5+4\omega}b^{-2}e^{-2i\vartheta}\phi_{0}, \qquad  
     \tilde{\Psi}_{1} =   \Xi^{4+2\omega}b^{-1}e^{-i\vartheta}\phi_{1}, \qquad 
     \tilde{\Psi}_{2} = \Xi^{3}\phi_{2},  \nn \\ 
     & \tilde{\Psi}_{3} = \Xi^{2-2\omega}b\, e^{i\vartheta}\phi_{3}, \qquad\quad\;\; 
     \tilde{\Psi}_{4} =   \Xi^{1-4\omega}b^2 e^{2i\vartheta}\phi_{4}. \label{eq:PhysicalWeyl-ToRescaledWeyl}
\end{eqnarray}
A natural choice for $\omega$ made in the classical theory of asymptotics is $\omega=0$ from which the peeling theorem follows if one assumes that the componentes of rescaled Weyl tensor in the unphysical basis
are finite at $\scri^+$ and the boost funciton is of order $b = \mathcal{O}(1)$. In that case, for instance, Bondi coordinates are usually chosen so that $\Xi \simeq 1/r$ --- see \cite{PenRin84, Val16}. In terms of the unphysical null tetrads \eqref{tetrad_trans_general}, the choice $\omega = 0$ implies that the outgoing null vector $\ell^a$ incorporates entirely the effect of the conformal re-scaling, whereas the unphysical ingoing null vector $k^a$ differs from its physical counterpart at most by a boost transformation.

\smallskip
As it is clear from equation~\eqref{eq:PhysicalWeyl-ToRescaledWeyl} the Weyl scalars transform homogeneously under the conformal re-scaling \eqref{eq:ConformalTransformation}. Quantities transforming homogeneously under conformal transformations are called  \emph{conformal densities}. 

\smallskip
This property is particular important to the Petrov classification. In particular, if one considers a spacetime  $(\tilde{\mathcal{M}},\tilde{\bmg})$ of Petrov-type D, i.e., admitting two pairs of principal null directions (PNDs), see \cite{PenRin84}, then $\tilde{\Psi}_{2}$ is the only non-zero independent component of the Weyl tensor if we take the physical null tetrad to be aligned with the PNDs.  

\smallskip
Therefore, it is clear from the definition \eqref{eq:d} that the only non-zero independent component of the rescaled Weyl tensor is encoded in $\phi_2$, if the unphysical null tetrad  $\{l^a, k^a, m^a, m^{*a}\}$ is also aligned with the principal null directions as in equations~\eqref{tetrad_trans_general}, i.e. if no null-rotation is performed.

\smallskip

The assumptions of a Petrov D background, also implies the vanishing of the spin-coefficients $(\tilde \kappa, \tilde \sigma, \tilde \nu, \tilde \lambda)$ via the Goldberg-Sachs theorem \cite{Gold62}, which in turn plays an important role in the original derivation of the Teukolsky equation~\cite{Teu73}. Tables (4.5.29) and (5.6.28) of Ref.~\cite{PenRin84} brings a full list of conformal transformation laws for the spin-coefficients, and one observes that  $(\kappa, \sigma, \nu, \lambda)$ are also conformal densities. Thus, we obtain the following proposition

\begin{proposition}\label{prop:petrovtypeD}
  If $(\mathring{\tilde{\mathcal{M}}},\mathring{\tilde{\bmg}})$,
  satisfies
  \begin{eqnarray}\label{eq:WeylToRescaledWeyl}
   & \mathring{\tilde{\Psi}}_{0}= \mathring{\tilde{\Psi}}_{1} = \mathring{\tilde{\Psi}}_{3} =
    \mathring{\tilde{\Psi}}_{4}=0. \noindent \\ & \mathring{\tilde{\kappa}}=
    \mathring{\tilde{\sigma}}= \mathring{\tilde{\nu}}=
    \mathring{\tilde{\lambda}}=0.
  \end{eqnarray}
  then $(\mathring{\mathcal{M}},\mathring{\bmg}, \mathring{\Xi})$
  where $\mathring{\bmg}= \mathring{\Xi}^2\tilde{\bmg}$, satisfies
  \begin{eqnarray}
   & \mathring{\phi}_{0}= \mathring{\phi}_{1} = \mathring{\phi}_{3} =
    \mathring{\phi}_{4}=0. \noindent \\ & \mathring{{\kappa}}=
    \mathring{{\sigma}}= \mathring{{\nu}}= \mathring{{\lambda}}=0.
  \end{eqnarray}
  where the components $\phi_n$  of the 
   the rescaled Weyl tensor and the spin-coefficients are computed respect to a $\mathring{\bmg}-$normalised null tetrad adapted to the PNDs
\end{proposition}

However, the remaining spin-coefficients ($\epsilon, \alpha, \beta, \gamma, \pi,\mu,\rho,\tau$) are not conformal densities. As a consequence, it is not generally clear whether the vanishing or specific combinations of physical spin-coefficients (such as those appearing in the original Teukolsky equation~\cite{Teu73}) would translate into analogous statements in the unphysical spacetime (but see \cite{Schneider:2025oqq} for a conformal extension of the Geroch-Held-Penrose formalism). 

\smallskip

In the context of the Teukolsky equation, the spin-coefficient $\tilde \rho$ and the Killing-Yano coefficient $\tilde \zeta$ play an important role. The former is a spin-coefficient defined via
\begin{equation}
    \tilde \rho = - \tilde m^a \tilde m^{\ast}{}^{b} \, \tilde \nabla_a \tilde \ell_b,
\end{equation}
whereas the latter is defined in terms of the conformal Killing-Yano tensor
\begin{equation}
   \tilde f_{ab} = (\tilde \zeta + \tilde \zeta^\ast) \tilde k_{[a} \tilde \ell_{b]} - (\tilde \zeta - \tilde \zeta^\ast) \tilde m^\ast{}_{[a} \tilde m_{b]}. 
\end{equation}
In a Petrov-type D spacetime, the Killing-Yano coefficient $\tilde{\zeta}$ is directly related to Weyl scalar $\tilde \zeta \propto \tilde \Psi_2^{1/3}$  spacetime and it relates to the expansion via
\begin{equation}
\label{eq:rho_zeta}
\dfrac{\tilde \rho}{\tilde \rho^*} =  \dfrac{\tilde \zeta^\ast}{\tilde \zeta}
\end{equation}
Incidentally, in the case of the Kerr spacetime, with PNDs given by the Kinnersley tetrad, their inverse coincide ($\tilde{\zeta}=\tilde{\rho}^{-1}$) and this leads in some accounts to them being taken interchangeably in the usual literature about the Teukolsky equation, Indeed, it is common to find definitions of the $-2$ spin-weighted Teuksolsky master function ${}_{-2} \tilde \Psi $ expressed in terms of either quantities
\begin{equation}
\label{eq:Psi_minus2}
{}_{-2} \tilde \Psi = \tilde \rho^{-4} \check{\tilde  {\Psi}}_4 = \tilde \zeta^{4} \check{\tilde {\Psi}}_4.
\end{equation}

When mapping into the unphysical spacetime, however, the expansion $\tilde \rho$ {\em is not} a conformal density, and as we will show, it may even vanish in some cases. In contrast, the  Killing-Yano coefficient is a conformal density satisfying
\beq
\label{eq:KY_conf}
\tilde \zeta = \Xi^{-1}  \zeta
\eeq
which follows from the conformal transformation law for the Killing-Yano tensor. The meaning of $\tilde{\zeta}$ and  above discussion is particularly transparent in spinor notation since for Petrov-type D backgrounds $\tilde{\zeta}_{AB}=\tilde{\zeta}\tilde{o}_{(A}\tilde{\iota}_{B)}$ is a Killing spinor (i.e. a spinor satisying the $\tilde{\nabla}^{A'}{}_{(A}\tilde{\zeta}_{BC)}=0$) if and only if $\tilde{\zeta}=q\tilde{\Psi}_{2}^{-1/3}$ where $q$ is a proportionality constant ---see for instance \cite{Buc58,And15, GasWil22}. However, we will avoid using spinor notation in this article and follow the tensorial definitions for $\tilde{\zeta}$ given in ~\cite{PouWar21} and employ $\tilde \zeta$ as the core geometrical object defining the \emph{master variable} \eqref{eq:Psi_minus2}.
In the case of the unphysical spacetime
the the unphysical Killing-Yano coefficient $\zeta$ corresponds to the only non-zero component of the rescaled Weyl tensor: 
\begin{equation}
\label{Killing_Spinor_Factor_DefsMaster}
\zeta = q\mathring{\phi}_{2}^{-1/3}.
\end{equation}
where $q$ is a proportionality constant which will not play a role in our discussion.
\smallskip

With these elements, we discuss in the next section, a derivation of the conformal Teukolsky directly from CEFEs which bypasses the complexity of the conformal transformation formulae for the spin-coefficients altogether by working directly at the level of the unphysical spacetime quantities. 
\subsection{The CEFEs in the Newman-Penrose notation}

As discussed earlier, the CEFEs \eqref{CFE_tensor_zeroquants_XiEq}-\eqref{standardCEFERescWeyl} are geometric relations for the curvature tensors of the unphysical spacetime. From these, one can construct various hyperbolic reductions and PDE formulations of the same underlying tensorial equation. In the previous section, we outlined a metric-based, second-order formulation of the CEFEs. However, to derive the CEFE-Teukolsky equation, it is necessary to employ a non-metric hyperbolic reduction that incorporates the Newman–Penrose language directly into the conformal spacetime. This formulation was developed in \cite{HilValZha20}, and we will not reproduce the full derivation here; instead, we present only the schematic structure of the relevant equations. For the detailed expressions, the reader is referred to Appendix A of \cite{HilValZha20}.

Let $\bme$, $\bm\Gamma$, $\bm\Phi$ and $R$ schematically encode the null tetrad, the connection coefficients, the tetrad components of the trace-free Ricci and Ricci scalar of the unphysical spacetime $(\mathcal{M},\bmg)$. 
Similarly, let $\bm\nabla$ schematically denote directional derivatives along $\bme$ --- see table \ref{Tab:NPformalism}. Additionally, let $\Xi$, $s$  $\bm\phi$ denote the conformal factor, the Friedrich scalar and  the tetrad components of the rescaled Weyl tensor respectively. 

Then, the NP-formulation of the CEFEs has the schematic form:

\label{eq:schematicNPCEFEs}
    \begin{eqnarray}
    &\bm\nabla \bm\nabla \Xi  \simeq\bm\nabla\Xi * \bm\Gamma + \Xi * \bm\Phi  + \Xi R+ s\label{nablanablaXi}\\ 
    &\bm \nabla s \simeq R*\bm\nabla\Xi + \bm\Phi*\bm\nabla\Xi \label{nablaS}\\ 
    &\bm\nabla \bm\Phi + \bm\nabla R  \simeq \bm\Phi * \bm\Gamma + \bm\Gamma*\bm\phi
    \label{eq:bianchicefeRicci}\\
    & \bm\nabla \bm\phi \simeq \bm\Gamma*\bm\phi \label{eq:bianchicefeWeyl}\\
    & \bm\nabla\bm\Gamma \simeq R +\bm\Phi + \Xi\bm\phi + \bm\Gamma*\bm\Gamma \label{eq:cartan1}\\
    & \bm \nabla \bme \simeq  \bm\Gamma \label{eq:cartan2}
\end{eqnarray}
with the multiplicative operator $\ast$ in expressions $X * Y$ representing a linear combination of some components of $X$ by some other components of $Y$. 
Observe that \eqref{eq:cartan1}-\eqref{eq:cartan2} are just the \emph{Cartan structure equations} where the conformal factor $\Xi$ enters since the Weyl tensor $\bm\Psi$ and the rescaled Weyl tensor $\bm\phi$ are related as $\bm\Psi = \Xi \bm\phi$. Also, observe that
equations \eqref{eq:bianchicefeRicci} and \eqref{eq:bianchicefeWeyl} correspond
to equations \eqref{standardCEFESchouten} and \eqref{standardCEFERescWeyl} which are the \emph{Bianchi sector} of the CEFEs.
Similarly, equations \eqref{nablanablaXi} and \eqref{nablaS} encode \eqref{StandardCEFEsecondderivativeCF} and \eqref{standardCEFEs} governing the evolution of the conformal factor $\Xi$ and the Friedrich scalar $s$.
\medskip

\begin{table}[h]
\centering
\begin{tabular}{|l|l|}
\hline
\hline
Tetrad & $\bme:=\{\ell^a,  k^a,  m^a, m^{*a} \}$ \\
\hline
Directional derivatives & $\bm\nabla = \{D=\ell^a\nabla_a,\;\Delta =k^a\nabla_a\;, \delta = m^a\nabla_a,\; \delta^* =  m^{*a}\nabla_a \}$ \\
\hline
Connection (spin) coefficients & $\bm\Gamma = \{\kappa,\;\tau,\;\sigma,\;\rho,\epsilon,\;\gamma,\;\beta,\;\alpha,\; \pi,\;\nu,\;\mu,\; \lambda\}$ \\
\hline
Rescaled Weyl tensor & $\bm\phi = \{\phi_0,\;\phi_1,\;\phi_2,\;\phi_3,\;\phi_4\}$ \\
\hline
Trace-free Ricci tensor & $\bm\Phi= \{\Phi_{00},\;\Phi_{01},\;\Phi_{02},\;\Phi_{12},\;\Phi_{22} \}$ \\
\hline
\end{tabular}
\caption{Notation for the Newman-Penrose formulation of  CEFEs.}
\label{Tab:NPformalism}
\end{table}

\subsection{Conformal Teukolsky equations}

To derive the conformal Teukolsky equation implied
by the CEFEs, inspired by the analogous classical derivation of the physical
Teukolsky equation in \cite{Teu73}, one starts by identifying to the following subset of equations in \eqref{eq:bianchicefeWeyl} and
\eqref{eq:cartan2}:
\label{eq:startingpointTeuk}
\begin{eqnarray}
  & (\delta^*-4\alpha + \pi)\phi_0 - (D - 4 \rho - 2
  \epsilon)\phi_1 -3 \kappa \phi_2 =0, \label{eq:Bianchi1} \\ & (\Delta
  -4 \gamma + \mu) \phi_0 - (\delta - 4 \tau -2 \beta)\phi_{1} - 3
  \sigma \phi_2 =0, \label{eq:Bianchi2} \\ &
 (D - \rho - \rho^* -3 \epsilon +
  \epsilon^*)\sigma - (\delta -\tau + \pi^*-\alpha^*
  -3\beta)\kappa =   \Xi\phi_0 . \label{eq:cartanstruct-rescweyl}
\end{eqnarray}
These correspond to equations (21e), (21a) and (13f) of Appendix A in
\cite{HilValZha20}. 

Observe that equations \eqref{eq:Bianchi1}-\eqref{eq:cartanstruct-rescweyl} hold independently of the choice of $\omega$ or the spin-boost parameters $(b,\vartheta)$, as their derivation does not originate from any conformal rescaling of the physical Newman–Penrose equations. Instead, it is carried out entirely within the framework of the CEFEs, which are formulated solely in terms of unphysical fields, making no reference to the physical spacetime or its associated field equations.
Notice in particular that equation \eqref{eq:cartanstruct-rescweyl} illustrates how, in general, the conformal factor $\Xi$ couples in the CEFEs. Recall that, in the CEFEs, the conformal factor satisfies its own system of evolution equations, which—in Newman–Penrose notation—are given schematically by equation ~\eqref{nablanablaXi}, and explicitly as equations (15a)–(15j) in \cite{HilValZha20}.

\medskip

By linearising equations \eqref{eq:Bianchi1}, \eqref{eq:Bianchi2}, and \eqref{eq:cartanstruct-rescweyl}—that is, expanding the fields as $Q = \mathring{Q} + \check{Q}$ and retaining only linear terms and using Proposition \ref{prop:petrovtypeD}, one obtains:
\label{eq:LinearSecondStepTeuk}
\begin{eqnarray}
  & (\mathring{\delta}^*-4\mathring{\alpha} +
  \mathring{\pi})\check{\phi}_0 - (\mathring{D} - 4\mathring{\rho} - 2
  \mathring{\epsilon})\check{\phi}_1 -3 \check{\kappa}
  \mathring{\phi}_2 =0, \label{eq:Bianchi1-lin} \\ &
  (\mathring{\Delta} -4 \mathring{\gamma} + \mathring{\mu})
  \check{\phi}_0 - (\mathring{\delta} - 4\mathring{\tau} -2
  \mathring{\beta})\check{\phi}_{1} - 3\check{\sigma}\mathring{\phi}_2
  =0, \label{eq:Bianchi2-lin} \\ & (\mathring{D} - \mathring{\rho} -
  \mathring{\rho}^* -3\mathring{\epsilon} +
  \mathring{\epsilon}^*)\check{\sigma} - (\mathring{\delta}
  -\mathring{\tau} + \mathring{\pi}^*-\mathring{\alpha}^*
  -3\mathring{\beta})\check{\kappa} =
  \mathring{\Xi}\check{\phi}_0.\label{eq:cartanstruct-rescweyl-lin}
\end{eqnarray}
\begin{remark}
\emph{
Note that it is not a priori obvious that the perturbation of the conformal factor, $\check{\Xi}$, does not enter the equations—see, for instance, equation \eqref{eq:weyltensor}. However, the absence of the linear term $\mathring{\phi}_0 \check{\Xi}$ in equation \eqref{eq:cartanstruct-rescweyl-lin} is a consequence of linearising around a Petrov type D background, which has $\mathring{\phi}_0 = 0$.
}
\end{remark}

After decoupling $\check{\phi}_0$ from equations \eqref{eq:Bianchi1-lin}, \eqref{eq:Bianchi2-lin}, and \eqref{eq:cartanstruct-rescweyl-lin}, the subsequent derivation of the CEFE-Teukolsky equation resembles, in structure, the classical argument leading to the physical Teukolsky equation in \cite{Teu73}. However, the conformal setting introduces an important subtlety: the presence of the conformal factor and its derivatives in the CEFEs could, in principle, spoil the decoupling mechanism. For this reason, we retrace the derivation step by step within the conformal formalism, explicitly verifying that no additional $\mathring{\Xi}$ or $\check{\Xi}$ contributions enter the final equation. This check is essential to establish that the CEFE-Teukolsky equation genuinely inherits the decoupling property in the conformal framework. Since $(\mathring{\mathcal{M}},\mathring{\bmg},\mathring{\Xi})$
is a Petrov-type D solution to the CEFEs, using equation (21f) and (21b)
from \cite{HilValZha20} one has
\begin{eqnarray}\label{eq:21f21bPetrovD}
  \mathring{D}\mathring{\phi}_2=3\mathring{\rho}\mathring{\phi}_2, \qquad
  \mathring{\delta}\mathring{\phi}_2=3\mathring{\tau}\mathring{\phi}_2.
\end{eqnarray}
Multiplying equation \eqref{eq:cartanstruct-rescweyl-lin}
by $\mathring{\phi}_2$ and using equation \eqref{eq:21f21bPetrovD} one gets
\begin{eqnarray}
(\mathring{D} - 4\mathring{\rho} -
  \mathring{\rho}^* -3\mathring{\epsilon} +
  \mathring{\epsilon}^*)(\check{\sigma}\mathring{\phi}_2) - (\mathring{\delta}
  -4\mathring{\tau} + \mathring{\pi}^*-\mathring{\alpha}
  -3\mathring{\beta})(\check{\kappa}\mathring{\phi}_2) =
  \mathring{\Xi}\mathring{\phi}_2\check{\phi}_0.\label{eq:cartanstruct-rescweyl-lin-refined}
\end{eqnarray}
At this point one needs to make use of the following commutator
which holds for any two real constants $p$ and $q$:
\begin{eqnarray}\label{eq:commutator}
 & (\mathring{D} -(p+1)\mathring{\epsilon} + \mathring{\epsilon}^*
  + q\mathring{\rho} - \mathring{\rho}^* )
  (\mathring{\delta}-p\mathring{\beta}+q\mathring{\tau})
  \nonumber \\ & \qquad \qquad \qquad  -(\mathring{\delta} - (p+1)\mathring{\beta}
  -\mathring{\alpha}^*+ \mathring{\pi}^* + q\mathring{\tau})
  (\mathring{D}-p\mathring{\epsilon} + q\mathring{\rho}) = 0.
\end{eqnarray}
  The last commutator follows from equations (12), (13b), (13e), (13q)  in Appendix A of
\cite{HilValZha20} and Proposition \ref{prop:petrovtypeD}.
  Applying $(\mathring{D}
-3\mathring{\epsilon} +\mathring{\bar{\epsilon}} - 4\mathring{\rho} -
\mathring{\bar{\rho}})$ to equation \eqref{eq:Bianchi2-lin} and
$(\mathring{\delta} + \mathring{\bar{\pi}}-\mathring{\bar{\alpha}}
-3\mathring{\beta} -4\mathring{\tau})$ to equation
\eqref{eq:Bianchi1-lin}, substracting the two, using
the commutator \eqref{eq:commutator} and equation \eqref{eq:Bianchi2-lin}, a long but direct calculation
gives 
\begin{eqnarray}\label{eq:conformal_teukolsky}
 [ (\mathring{D}
&-3\mathring{\epsilon} +\mathring{\epsilon}^* - 4\mathring{\rho} -
\mathring{\rho}^*)(\mathring{\Delta}-4\mathring{\gamma} + \mathring{\mu})
\\ & -(\mathring{\delta} + \mathring{\pi}^*-\mathring{\alpha}^*
-3\mathring{\beta} -4\mathring{\tau})  (\mathring{\delta}^* + \mathring{\pi} -4\mathring{\alpha}) \nonumber- 3\mathring{\Xi}\mathring{\phi}_2]\check{\phi}_0=0.
\end{eqnarray}
Exploiting the $\l^a \leftrightarrow k^a$ and 
$m^a \leftrightarrow m^{*a}$ symmetry one obtains the analogous equation for $\check{\phi}_4$.
This discussion is summarised in the following:

\begin{proposition}\label{prop:CEFETeukolsky}
  Let $(\mathring{\mathcal{M}},\mathring{\bmg}, \mathring{\Xi})$ denote a Petrov-type D solution to the (non-linear) CEFEs.
  The linearisation of the CEFEs around
  $(\mathring{\mathcal{M}},\mathring{\bmg}, \mathring{\Xi})$ imply the \emph{CEFE-Teukolsky} equation:
\label{eq:conformal_teukolsky}
\begin{eqnarray}\label{eq:conformal_teukolsky_phi0}
 [ (\mathring{D}
&-3\mathring{\epsilon} +\mathring{\epsilon}^* - 4\mathring{\rho} -
\mathring{\rho}^*)(\mathring{\Delta}-4\mathring{\gamma} + \mathring{\mu}) 
 \\ & -(\mathring{\delta} + \mathring{\pi}^*-\mathring{\alpha}^*-3\mathring{\beta}   -4\mathring{\tau}  )(\mathring{\delta}^* + \mathring{\pi} -4\mathring{\alpha}) \nonumber- 3\mathring{\Xi}\mathring{\phi}_2]\check{\phi}_0=0.
\end{eqnarray}
\begin{eqnarray}
\label{eq:conformal_teukolsky_phi4}
 [ (\mathring{\Delta}
&+3\mathring{\gamma} -\mathring{\gamma}^* + 4\mathring{\mu} +
\mathring{\mu}^*)(\mathring{D}+4\mathring{\epsilon} - \mathring{\rho})
\\ &-(\mathring{\delta}^* - \mathring{\tau}^*+\mathring{\beta}^*
+3\mathring{\alpha} +4  \mathring{\pi} )(\mathring{\delta}^* - \mathring{\tau} +4\mathring{\beta}) \nonumber - 3\mathring{\Xi}\mathring{\phi}_2]\check{\phi}_4=0.
\end{eqnarray}
Here the spin-coefficients and directional derivatives of the Newman-Penrose formalism are computed with respect to the unphysical tetrad normalised with respect to $\mathring{\bmg}$ and aligned to the PNDs. Similarly $\mathring{\phi}_{2}$, $\check{\phi}_{0}$ and $\check{\phi}_{4}$ refer to the background and perturbation of the components of the rescaled Weyl tensor in
the unphysical tetrad. These quantities written in terms of the background (ring-quantities) and perturbation (check-quantities) Weyl scalars computed in the unphysical and physical tetrad (tilde-quantities), respectively, read:
\begin{eqnarray}
 & \mathring{\phi}_2= \mathring{\Xi}^{-1}\mathring{\Psi}_2= \mathring{\Xi}^{-3}\mathring{\tilde{\Psi}}_{2},
\\
& \check{\phi}_0=\mathring{\Xi}^{-1}\check{\Psi}_0 =
\mathring{\Xi}^{-5-4\omega}b^4e^{2i\vartheta}\check{\tilde{\Psi}}_{0}, \\
&
\check{\phi}_4=\mathring{\Xi}^{-1}\check{\Psi}_4 =
\mathring{\Xi}^{-1+4\omega}b^{-4}e^{-2i\vartheta}\check{\tilde{\Psi}}_{0},
\end{eqnarray}
where $(b,\vartheta)$ represents a general spin-boost of the physical null tetrad.
\end{proposition}
As a final step, we introduce the conformal master functions  with spin weight $\mathfrak{s}$ denoted  as ${}_\mathfrak{s} \phi$ and defined by
\begin{eqnarray}
\label{eq:conformal_MasterFunc}
    {}_{\mathfrak{s} }\phi = \left\{
\begin{array}{ccc}
\check{\phi}_{0} & \rm{if} & \mathfrak{s} =2, \\
\mathring\zeta^4 \check{\phi}_{4}   & \rm{if} & \mathfrak{s} =-2
\end{array}
\right.
\end{eqnarray}
with $\mathring\zeta$ the conformal Killing-Yano coefficient \eqref{Killing_Spinor_Factor_DefsMaster} associated to the conformal background spacetime.

\section{Applications}\label{sec:Results}
We now turn to concrete applications of the formalism developed in the previous sections, focusing on the derivation of the conformal Teukolsky equation in different conformal representations of the Kerr spacetime. Our first examples explore representations arising naturally from hyperboloidal coordinate systems, providing a geometric interpretation of results already present in the literature. As a second case study, we extend the formalism to construct, for the first time, a representation of the Teukolsky equation adapted to a conformal framework where spatial infinity is realized as a cylinder.

\smallskip
We begin with a brief review of the Kerr solution in standard Boyer–Lindquist (BL) coordinates, which will establish the notation used throughout this section. In standard BL coordinates $\tilde{x}^\mu = (t, r, \theta, \varphi)$, the Kerr spacetime is described by the metric
\begin{eqnarray}
\label{KerrMetricBL}
\tilde{g}_{\mu\nu} d\tilde{x}^\mu d \tilde{x}^\nu  &=& -\tilde f\, dt^2 - \frac{4Mar}{\tilde\Sigma} \sin^2\theta\, dt\, d\phi + \frac{\tilde\Sigma}{\tilde\Delta}\, dr^2 + \tilde\Sigma\, d\theta^2 \\
&& + \sin^2\theta \left(\tilde\Sigma_0  + \frac{2Ma^2 r}{\tilde\Sigma} \sin^2\theta \right) d\phi^2,   \nn
\end{eqnarray}
where the auxiliary functions take the form
\begin{eqnarray}
\label{eq:Delta}
\tilde\Delta &= r^2 - 2Mr + a^2 = (r - r_+)(r - r_-), \qquad \tilde f = 1 - \frac{2Mr}{\Sigma}, \\
\label{eq:Sigmas}
\tilde\Sigma &= r^2 + a^2 \cos^2\theta, \qquad \tilde\Sigma_0 = \left.\tilde\Sigma\right|_{\theta=0} = r^2 + a^2.
\end{eqnarray}

Here, $M$ and $a$ denote the black hole’s mass and angular momentum, respectively. The roots of $\tilde{\Delta}$ correspond to the Cauchy horizon $r_-$ and the event horizon $r_+$, given explicitly by
\begin{equation}
\frac{r_\pm}{M} = 1 \pm \sqrt{1 - \frac{a^2}{M^2}}.
\end{equation}

In our analysis, it is convenient to adopt a rescaled parametrisation in which the event horizon radius $r_+$ sets the natural length scale. This leads to a dimensionless rotation parameter defined by
\begin{equation}
\kappa = \frac{a}{r_+} = \frac{a}{M + \sqrt{M^2 - a^2}} \quad \Longleftrightarrow \quad \kappa^2 = \frac{r_-}{r_+}.
\end{equation}
With this choice, the mass parameter is re-expressed as
\begin{equation}
M = \frac{r_+}{2}(1 + \kappa^2).
\end{equation}
Further important quantities for the Kerr metric are the horizons angular velocities $\Omega$ and surface gravities $\varkappa$, respectively 
\begin{eqnarray}
& \Omega_{+} = \dfrac{a}{2M \rh} = \dfrac{\kappa}{\rh (1+\kappa^2)}, \quad \varkappa_{+} = \dfrac{1}{4M} - M \Omega_{+} ^2 = \dfrac{1-\kappa^2}{2\rh (1+\kappa^2)}   \\
& \Omega_{-} = \dfrac{a}{2M \rc} = \dfrac{\kappa^{-1}}{\rh  (1+\kappa^2)}, \quad \varkappa_{-} = \dfrac{1}{4M} - M \Omega_{-} ^2 = - \dfrac{\kappa^{-2}\left(1-\kappa^2\right)}{2\rh  (1+\kappa^2)}.
\end{eqnarray}

The choice of a tetrad aligned with the PNDs plays a central role in the derivation of the Teukolsky equation. The Kinnersley tetrad $\{ \tilde \ell^a,  \tilde k^a,  \tilde m^a,  \tilde m^{\ast}{}^a  \}$ was employed in the original work~\cite{Teu73}. In terms of BL coordinates $\tilde{x}^\mu$, their components read
\begin{eqnarray}
\label{eq:Kin_null_tetrad_l}
&\tilde \ell^{\tilde \mu} =\bigg( \dfrac{r^2+a^2}{\Delta(r)}, 1,0, \dfrac{a}{\Delta(r)}\bigg), \\ 
\label{eq:Kin_null_tetrad_k}
&\tilde k^{\tilde \mu} =\dfrac{1}{2\Sigma(r, \theta)}\bigg( \left(r^2+a^2\right), -\Delta(r) ,0, a \bigg), \\
\label{eq:Kin_null_tetrad_m}
& \tilde m^{\tilde \mu} = \dfrac{1}{{\sqrt{2} (r+ i a \cos\theta)}} \left( i a \sin\theta, 0,1 , \dfrac{i}{\sin\theta}\ \right).
\end{eqnarray}
%
Moreover, the non-vanishing Weyl scalar $\tilde \Psi_2$ and the associated Killing-Yano coefficient $\tilde \zeta = - M^{1/3} \tilde \Psi_2^{-1/3}$ read
\beq
\tilde \Psi_2 = - \dfrac{M}{(r- i a \cos\theta)^3}, \quad \tilde \zeta  = r- i a \cos\theta.
\eeq

In the next sections we will introduce scenarios for which the Kerr line-element \eqref{KerrMetricBL} naturally undergo a conformal re-scaling $g_{ab} = \Xi^2 \tilde g_{ab}$. With respect to the conformal metric, it is convenient to defined the conformal metric functions~\cite{Mac20} 
\begin{equation}
\label{eq:conf_metric_func}
\Delta = \Xi^2 \tilde \Delta, \quad \Sigma = \Xi^2 \tilde \Sigma, \quad \Sigma_0 = \Xi^2 \tilde \Sigma_0.
\end{equation}

\subsection{Hyperboloidal framework}
The transformation of the standard Boyer–Lindquist coordinates $\tilde{x}^\mu$ into a hyperboloidal coordinate system $x^\mu=\{T, \varsigma, \theta, \varphi\}$ in the so-called minimal gauge class~\cite{Mac20} follows from
\begin{equation}
\label{eq:hyp_transfo}
t = \lambda \bigg( T - H\left(\varsigma\right) \bigg), \quad r = r_o + \dfrac{\rh-r_o}{\varsigma}, \quad \phi = \varphi - \varphi_*(\varsigma)
\end{equation}
with the height function given by
\begin{equation}
H(\varsigma) = - \dfrac{\rh-r_o}{\lambda\, \varsigma}  + \dfrac{2M}{\lambda} \ln \varsigma + \dfrac{1}{2 \lambda \varkappa_+ } \ln(1- \varsigma) + \dfrac{1}{2 \lambda \varkappa_- } \ln\left(1- \dfrac{\varsigma}{\varsigma_-}\right),
\end{equation}
and the angular ``tortoise" coordinate defined by
\beq
\varphi_*(\varsigma) = \phi_*(r(\varsigma)), \quad \dfrac{d }{dr} \phi_*(r) = \dfrac{a}{\tilde \Delta}.
\eeq
In the above expressions, $\lambda$ is a given length scale of the spacetime that will be adjusted to match definitions in different works. The radial transformation maps $r\to \infty$ into $\varsigma = 0$, and it conveniently sets the black hole horizon $\rh$ to $\varsigma_+ = 1$. The Cauchy horizon $\rc$ is mapped into $\varsigma_- = \dfrac{\rh - r_o}{\rc - r_o}$, and it is, therefore, controlled by the parameter $r_o$. The simple choice $r_o = 0$ is dubbed as \emph{the radial fixing gauge} and it allows $\varsigma_- = \kappa^{-2}$ to depend on the black hole rotation parameter. The Cauchy horizon fixing gauge fixes $\varsigma_- \to \infty$ with the choice $r_o = \rc$. Further details on the proprieties of these choices can be found in~\cite{Mac20, Minucci:2024qrn, Assaad:2025nbv}.

\smallskip
The most relevant elements for the construction of the conformal Teulkolsky equation are the specification of the metric's conformal mapping $g_{ab} = \Xi^{2} \tilde g_{ab}$, and a choice of conformal tetrad $\{ \ell^a,  k^a,  m^a,  m^{\ast}{}^a  \}$. 

\smallskip
Within the strategy followed in Ref.~\cite{Mac20,PanossoMacedo:2023qzp} for the hyperboloidal framework, the compact radial coordinate $\varsigma$ is identified directly with the conformal factor via $\Xi = \varsigma/\lambda$. Then, the conformal null tetrad basis  $( \ell^a,  k^a,  m^a, m^\star{}^a)$ associated with the conformal metric $ g^{ab} = -2  \ell^{\,(a}  \ell^{\,b)}  + 2  m^{\,(a}  m^\star{}^{\,b)}$ 
is obtained from the original Kinnersly tetrad basis via~\cite{Minucci:2024qrn}
\bea
\label{eq:conf_tetrad}
 \ell^a = \Xi^{-2} \dfrac{\Delta}{\lambda} \tilde \ell^a, \quad  k^a =  \dfrac{\lambda}{\Delta } \tilde k^a, \quad  m^a = \Xi^{-1}  \tilde m^a, \quad  m^\star{}^a = \Xi^{-1} \tilde m^\star{}^a.
\eea
A comparison against the general form of tetrad transformation \eqref{tetrad_trans_general} confirms the choice for the conformal weight $\omega=0$, naturally adapted to the peeling theorem at $\scri^+$ as demonstrated in equations~\eqref{eq:Pelling}. Besides, the boost factor reads $b=\dfrac{\Delta}{\lambda}$, with the length scale $\lambda$ ensuring the conformal tetrad is dimensionless. From this boost transformation, equations~\eqref{eq:Pelling} also indicate that the Teukolsky master function must re-scale as ${}_\s \tilde \Psi \sim {}_\s\phi \,\Delta^\s$. In previous works~\cite{Zenginoglu:2011jz,Mac20}, the pre-factor $\Delta^\s$ has always been included {\em ad hoc} to ensure a regular hyperboloidal master function. As in reference~\cite{Minucci:2024qrn}, this work emphasise the origin of the pre-factor as direct consequence of defining the Weyl scalar in terms of null tetrad most adapted to the hyperboloidal foliation. The above analysis provides us with an important conclusion:

\begin{remark} The hyperboloidal master function widely employed in the literature over the past decades~\cite{Zenginoglu:2011jz,Mac20} has the geometrical interpreation of being exactly the Newman-penrose components $\phi_n$ of the {\em re-scalled} Weyl tensor $d^a{}_{bcd}$, cf.~\eqref{unphysical_Weyl}.
\end{remark}

\smallskip
With the conformal structure within the hyperboloidal framework established, we procede to the explicit calculation of the conformal Teukolsky equation. For the purposes of this discussion, we focus on the radial fixing gauge --- cf.~\eqref{eq:hyp_transfo} with $r_o=0$, but the framework extend straighforwardly to   other choices of hyperboloidal foliation. 

The conformal metric functions \eqref{eq:conf_metric_func} read
\beq
\Delta =\left( \dfrac{\rh}{\lambda}\right)^2 (1-\varsigma)(1-\kappa^2 \varsigma), \quad \Sigma = \left( \dfrac{\rh}{\lambda}\right)^2 \left( 1+ \kappa^2 \varsigma^2 \cos^2\theta \right).
\eeq
whereas equations~\eqref{eq:Pelling} and \eqref{eq:KY_conf} yield the re-scaled Weyl scalar and conformal Killing-Yano coefficient in the background Kerr metric

\beq
\label{eq:phi2_hyp}
\phi_2 =-\dfrac{M}{\zeta^3}, \quad \zeta = \dfrac{\rh}{\lambda}\left( 1 - i \kappa \varsigma \cos\theta \right).
\eeq
Notice that we have simplified the notation by dropping all the mathring $\mathring{\cdot}$ in the background quantities. THat care is nevertheless needed to distinguish the background quantity $\phi_2$ from the perturbation master variable ${}_{\frak{s}}\phi$.

The components of conformal null tetrads \eqref{eq:conf_tetrad} adapted to the hyperboloidal foliation read explicitly
\label{eq:hyp_tetrad_radialfix}
\bea
\label{eq:hyp_tetrad_radialfix_l}
 \ell^\mu &=& \Big(  { \frac{2 \rh^2 (1 + \kappa ^2) (1 + \kappa ^2 (1 - \varsigma ))}{\lambda ^2}, -\frac{\lambda \Delta}{\rh },0,\frac{2 \rh \kappa }{\lambda }} \Big),  \\
\label{eq:hyp_tetrad_radialfix_k}
k^\mu &=&\dfrac{1}{\Sigma  } \Big( 1 + \varsigma (1 + \kappa ^2)  ,\frac{\lambda  \varsigma ^2}{2 \rh},0,0 \Big), \\
\label{eq:hyp_tetrad_radialfix_m}
m^\mu &=&\dfrac{1}{\sqrt{2}  \mathring{\zeta}^\ast} \Big(  \dfrac{ i \,\rh  \kappa  \sin\theta}{\lambda} ,0, 1,  \dfrac{i}{\sin\theta} \Big).  
\eea
Associated to this null tetrad one has 
the following non-zero spin coefficients:
\begin{eqnarray}\label{eq:UnphysicalKerrSpinCoeffs}
&\tau  = - \frac{i \sqrt{2} \kappa  \lambda  \sin\theta  \varsigma }{r_{+}{} (2 + \kappa ^2 \varsigma ^2 + \kappa ^2 \cos(2 \theta ) \varsigma ^2)}, \quad  
\epsilon  =  \dfrac{\rh \big( 1 +  \kappa ^2 (1  -2 \varsigma ) \big) }{2 \lambda}, 
\nn \\
&\beta  = \frac{\cot\theta }{2 \sqrt{2} {\zeta}^\star}, \quad
\alpha  = - \frac{r_{+}{} \csc\big(\dfrac{1}{2} \theta \big) \sec\big(\dfrac{1}{2} \theta \big) (2 \cos\theta  + i \kappa  ( \cos(2 \theta )-3) \varsigma )}{8 \sqrt{2} {\zeta} ^2 \lambda } \nn \\  
&\rho  = - \frac{i \Delta  \kappa  \cos\theta }{{\zeta} } \quad
\gamma  = - \frac{\varsigma }{2 {\zeta} ^2 {{\zeta}^\star }},  \quad
\pi  = \frac{i r_{+}{} \kappa  \sin\theta  \varsigma }{\sqrt{2} {\zeta}^2 \lambda } \quad
\mu  = - \frac{i \kappa  \cos\theta  \varsigma ^2}{2 {\zeta}^2 {\zeta }^\star}
\end{eqnarray}
Once again we draw the attention to the very important difference between the conformal spin-coefficient $\rho$ and the conformal Killing-Yano coefficient $\zeta$. Contrary to their behaviour in the physical spacetime, they are no longer trivially related to each other. In fact, the expansion $\rho$ actually vanishes in the Schwarzschild limit $\kappa = 0$, justifying the choice of $\zeta$ in the definition of the spin weighted conformal master function ${}_\s \phi$ in equation~\eqref{eq:conformal_MasterFunc}. Moreover, the Schwarzschild limit $\kappa = 0$ also shows a trivial relation for $\zeta$, i.e. with the conformal Killing-Yano coefficient assuming a constant value for $\kappa =0$, one has ${}_{-2} \phi$ being exactly $\phi_4$ in the Schwarzschild limit.

\smallskip
We finish this section with a direct calculation of the conformal Teukolsky equation, having references~\cite{AnsMac16} as benchmark for the Schwarzschild case ($\kappa=0$) and \cite{Mac20} for Kerr ($\kappa \neq 0$). To make the notation consistent with those works, we set $\lambda = 2\rh$ in the former and $\lambda = \rh$ in the latter. 

\medskip

For the Schwarzschild limit we perform a $1+1$ reduction via
\beq
{}_{\s} \phi (T, \varsigma,\theta, \varphi) = \sum_{\ell, m} {}_{\s} \phi_{\ell m} (T, \varsigma) {}_\s Y_{\ell, m}(\theta, \varphi),
\eeq
with ${}_\s Y_{\ell, m}(\theta, \varphi)$ the spin-weighted spherical harmonic. Then, proposition \ref{prop:CEFETeukolsky} yields
\begin{eqnarray}\label{Teukolsky_Mode_decomp_master}
&  [- (1 + \varsigma ) \partial _ {T}^2 +  (1 - 2 \varsigma ^2) \partial _ {T}\partial _
                {\varsigma} + \varsigma ^2 (1-\varsigma ) 
                \partial _ {\varsigma}^2 \nonumber \\ 
& + 
(s(\varsigma-1) - 2 \varsigma ) \partial _ {T} +  \varsigma
(s(2-\varsigma) +2 -3\varsigma) \partial _ {\varsigma}  \nn \\
& 
+ (-l(l+1)-(\varsigma-s)(1+s)]
  {}_{\s} \phi_{\ell m}=0,
  \end{eqnarray}
which coincides identically with that reported in \cite{AnsMac16}.

In the Kerr case, we further introduce a transformation in the angular coordinate $x=\cos\theta$ and we perform a $2+1$ reduction, with a regularisation at the symmetry axis $x=\pm 1$ via
\bea
\label{eq:2+1 decomp x}
{}_{\s} \phi (T, \varsigma,\theta, \varphi) = \sum_{m} {}_{\s} \phi_{m} (T, \varsigma, \theta) \left(  1+ x\right)^{\delta_1/2}  \left(  1- x \right)^{\delta_2/2} e^{i m \varphi},
\eea
 with $\delta_1 = |m-s|$ and $\delta_2 = |m+s|$. A calculation using proposition \ref{prop:CEFETeukolsky} and equation \eqref{eq:conformal_MasterFunc} gives
\begin{eqnarray}
\label{eq:HypTeuk}
&& 0  =  {}_{\s} \phi_{m}{}  \big[- (s - \dfrac{1}{2} ( \delta _{1}{} + \delta _{2}{})) 
(1 + s + \dfrac{1}{2} (\delta _{1}{} + \delta _{2}{}))
+ 2i m \kappa  \varsigma  + s (1 + \kappa ^2) \varsigma \nn\\ 
&&+ (1 + \kappa ^2 (1  -2 \varsigma )) \varsigma \big]  + 2 \big[2 (1 + \kappa ^2) (1 + \kappa ^2 (1  -2 \varsigma )) \varsigma 
 - \kappa ^2 \varsigma  (1 - (1 + \kappa ^2) \varsigma ) \nn \\ 
 &&
 + i m \kappa  (1 + 2 (1 + \kappa ^2) \varsigma ) 
 - s (-i \kappa  x 
 +    (1 +  \kappa ^2) (1  - (1 + \kappa ^2) \varsigma ))\big] \partial _{T}{}_{\s} \phi_{m}{} \nn \\ 
  && + \big[ \kappa ^2 ( x^2-1) + 4 (1 + \kappa ^2) (1 + \kappa ^2 (1  - \varsigma ))
 (1 + (1 +  \kappa ^2)  \varsigma )\big] \partial _{T}^2 {}_{\s} \phi_{m}{}  \nn \\ 
&&-2 \big[1 + \kappa ^2 \varsigma ^2  -2 (1 + \kappa ^2) (1 + \kappa ^2 (1  - \varsigma )) \varsigma ^2\big] \partial^2 _{T\varsigma}{}_{\s} \phi_{m}{} - \varsigma ^2(1  - \varsigma )  (1\!  - \! \kappa ^2 \varsigma )    \partial _{\varsigma}^2 {}_{\s} \phi_{m}{}  \nn   \\
&& - \varsigma  \big[2 (1 + s)  - \varsigma  (2i m \kappa  + (3 + s) (1 + \kappa ^2)   -4 \kappa ^2 \varsigma )\big] \partial _{\varsigma}{}_{\s} \phi_{m}{} \nn \\
&&   + ( \delta _{1}{} (x-1) + 2 x + \delta _{x}{} (1 + x)) \partial _{2}{}_{\s} \phi_{m}{} + (x^2-1) \partial _{x}^2 {}_{\s} \phi_{m}{}, \nn
\end{eqnarray}
which coincides identically with equation C1.1 of Appendix 1 in \cite{Mac20}.

We recall that equations~\eqref{Teukolsky_Mode_decomp_master} and \eqref{eq:HypTeuk} were originally derived in \cite{Mac20} via a direct change of coordinate and {\em ad hoc} inclusion of regularisation pre-factors. Here, they resulted from the rigours geometrical formulation of the conformal Teukolsky equation descending from the CEFEs. The expected consistency between the two approaches ensures the agreement of the framework developed in Sec.~\ref{sec:CEFE-Teukolsky} based on the CEFEs (when specialised to hyperboloidal coordinates) and the ad-hoc and coordinate-anchored approach in the past decades ~\cite{Zenginoglu:2011jz,Mac20}.  Having benchmarked the conformal Teukolsky equations, we proceed to apply the formalism to a different asymptotic region: the cylinder at spatial infinity.

\subsection{The cylinder at spatial infinity}

As discussed in the introduction, the conformal structure of asymptotically flat spacetimes with non-zero mass becomes degenerate at spatial infinity $i^0$ \cite{Pen65}. Various approaches address this issue through “blow-up’’ constructions that replace the point $i^0$ by an extended geometric structure \cite{Ger72,Fri98a,HinVas17}. A well-known example is Friedrich’s cylinder at spatial infinity \cite{Fri98a}, based on conformal geodesics and associated $F$-coordinates. However, explicit implementations of this construction remain technically challenging for rotating spacetimes such as Kerr.

For the Kerr spacetime, Ref.~\cite{Hennig:2020rns} introduced an alternative cylinder-like representation of $i^0$ using coordinates adapted to null geodesics. Although this construction does not correspond to Friedrich’s cylinder in a strict sense, it provides a practical framework for studying regular behaviour near the region where spatial and null infinity meet. In this section, we employ this representation to derive the CEFE-Teukolsky equation for spin weight $\s=\pm2$ on the conformally compactified Kerr spacetime.

\smallskip
The conformal representation of the Kerr spacetime described in Ref.~\cite{Hennig:2020rns} follows after a coordinate trasformation from the BL coordinates $\tilde x^\mu$ into the cylinder-like coordinates $x^\mu =\{ T, \varsigma, \theta, \upvarphi \}$ via,
\begin{numparts}
\bea
\label{eq:rCylider_trasfo}
&\tilde r = \dfrac{\rh}{\varsigma(1-T)}, \quad \phi = \upvarphi + \upvarphi_*(T,\varsigma) \quad {\rm with} \quad  \upvarphi_*(T,\varsigma) = \phi_*(r(T,\varsigma))\\
\label{eq:tCylider_trasfo}
&t = \rh \Bigg( \dfrac{T}{\varsigma(1-T)} - (1+\kappa^2)\ln(1-T) \nn \\ 
&+ \dfrac{1+\kappa^2}{1-\kappa^2} \Bigg[\kappa^2 \ln\left( \dfrac{1-\kappa^2 \varsigma}{1-\kappa^2 \varsigma(1-T)}   \right)  - \ln \left( \dfrac{1-\varsigma}{1-\varsigma(1-T)}\right) \Bigg] \Bigg).
\eea
\end{numparts}
The transformation \eqblock{eq:rCylider_trasfo}{eq:tCylider_trasfo} ensures that $i^0$ is mapped into $\varsigma=0$, whereas $T=1$ correspond to $\scri^+$. Besides, the time transformation \eqref{eq:tCylider_trasfo} is normalised such that the initial hypersurface $t=0$ corresponds to $T=0$.

As already pointed out, essential to the derivation of the conformal Teukolsky equation are the conformal factor $\Xi$ and the set of conformal null tetrads $\{\ell^a, k^a, m^a, m^\ast{}^a\}$ aligned with the PNDs. Here, they read~\cite{Hennig:2020rns}  $\Xi = 1/\tilde r = \varsigma(1-T)/\rh$ and the null tetrad adapted to the cylinder at $i^0$ follows from the Kinnersley \eqref{eq:Kin_null_tetrad_l}-\eqref{eq:Kin_null_tetrad_m} via
\bea
\label{eq:NullTetrad_ioBlow_trasfo}
& \ell^a =  \Xi^{-2} \dfrac{\varsigma}{\rh} \tilde \ell^a = \dfrac{\rh}{\varsigma(1-T)^2} \tilde \ell^a, \quad 
 k^a =  \dfrac{\rh}{\varsigma } \tilde k^a \\
& m^a = \Xi^{-1}  \tilde m^a, \quad  m^\star{}^a = \Xi^{-1} \tilde m^\star{}^a.
\eea
As in equation~\eqref{eq:conf_tetrad}, the above transformation incorporates the conformal re-scaling in the null tetrads with $\omega=0$, together with a boost $b=\dfrac{\varsigma}{\rh}$ according to the generic transformation \eqref{tetrad_trans_general}.

As in the hyperboloidal case, this conformal mapping naturally yields the peeling decay of the Weyl scalar at $\scri^+$. Indeed, for a constant $\varsigma \neq 0$, equations \eqref{eq:Pelling} leads to the decay in the physical Weyl components $\tilde \Psi_0 \sim (1-T)^5 \phi_0$ and $\tilde \Psi_4 \sim (1-T) \phi_4$ as $T \to 1$.

Here, however, in contrast with the hyperboloidal case \eqref{eq:conf_tetrad}, the boost factor also picks-up a factor $\varsigma$, which contributes to the asymptotic behaviour towards $i^0$. Thus, for a constant time slice $-1 < T < 1$, equations \eqref{eq:Pelling} provide both $\tilde \Psi_0 \sim \varsigma^3 \phi_0$ and $\tilde \Psi_4 \sim \varsigma^3 \phi_4$, as $\varsigma \to 0$.

Explicitly in terms of the coordinates $x^\mu=\{T, \varsigma, \theta, \upvarphi \}$, the null tetrad components read
\bea
\label{eq:NullTetrad_ioBlow_ell}
\ell^\mu &=&(1,0,0,0)  \\
k^\mu &=& \Bigg(\frac{(1-T)}{2 \Sigma}\left[ \frac{2\Sigma_0 F}{\varsigma^2} - (1-T)\Delta \right], \frac{F}{\varsigma} \frac{\Sigma_0}{\Sigma},0, \frac{\kappa \varsigma(1-T)^2}{\Sigma} \Bigg)\\
\label{eq:NullTetrad_ioBlow_m}
m^\mu&=&\frac{1}{\sqrt{2}(1+i \kappa \varsigma(1-T) \cos\theta)}
\Bigg(  \frac{F(1-T)}{\varsigma} i \kappa \sin\theta,  F \, i \kappa \sin\theta\, , 1  , \frac{i}{\sin\theta}  \Bigg). 
\eea
In the above expressions, the conformal metric functions are defined as in \eqref{eq:conf_metric_func}, and they depend on the coordinates $(T,\varsigma)$ via the radial transformation \eqref{eq:rCylider_trasfo}, i.e. $\Delta(T,\varsigma)$, $\Sigma(T,\varsigma)$ and $\Sigma_0(T,\varsigma)$. The function $F$, on the other hand, depends only on the radial coordinate $\varsigma$ and it generates the time transformation \eqref{eq:tCylider_trasfo} via the definition\cite{Hennig:2020rns}
\beq
F(\varsigma)=\dfrac{\varsigma^2 (1-\varsigma)(1-\kappa^2 \varsigma)}{1+\kappa^2 \varsigma^2} \Longrightarrow  \dfrac{t}{\rh} = \int_{\varsigma(1-T)}^\varsigma \dfrac{d\varsigma'}{F(\varsigma')}.
\eeq
A direct calculation renders:
\bea
\tau  &=& \frac{-i \kappa  \sin\theta  \varsigma  (1 - T )}{\sqrt{2} \Sigma}, \nn \\
\rho  &=&  \frac{-i \kappa  \cos\theta  \varsigma }{\zeta}, \nn \\
\gamma  &=&   \frac{i}{4\zeta \Sigma}(  - \kappa  \cos\theta  \varsigma ^2 (1 + \kappa ^2 (1 - 2 \varsigma  (1 - T ))) (1 - T )^2) (1 - T )  \nn\\&& + i (2 - \varsigma  (3 + \kappa ^2 (3 - 4 \varsigma  (1 - T ))) (1 - T )) 
  +\frac{F(\varsigma ) (1 + \kappa ^2 \varsigma ^2 (1 - T )^2)}{2 \varsigma ^2 \Sigma },\nn\\
\beta  &=&  \frac{i 2 \kappa  F(\varsigma ) \sin\theta   -2i \cot\theta  \varsigma }{ 4 \sqrt{2} \varsigma  \zeta }, \nn\\
\alpha  &=&  \frac{-i\kappa  F(\varsigma ) \sin\theta }{2 \sqrt{2} \varsigma  \zeta^\star} - \frac{4 \cot\theta   + 2i \kappa  (-3 + \cos(2 \theta )) \csc\theta  \varsigma  (1 - T )}{8 \sqrt{2} \zeta^{\star 2}},\nn\\
\pi  &=& \frac{i \kappa  \sin\theta  \varsigma  (1 - T )}{\sqrt{2} \zeta^2} ,\nn\\
\mu  &=& - \frac{\kappa  \cos\theta  \varsigma}{2 \zeta\Sigma}  (1 - \varsigma  (1 - T )) (1 - \kappa ^2 \varsigma  (1 - T )) (1 - T )^2,
\eea
where $\zeta$ is the Killing-Yano coefficient
which appears in the only non-zero component of the rescaled Weyl tensor:
\beq
\label{eq:phi2_i0}
\phi_2 =-\dfrac{M}{\zeta^3}, \quad \zeta = \left(1- i \kappa \cos\theta \varsigma (T-1)) \right)
\eeq
With all the above elements equations~\eqref{eq:conformal_teukolsky_phi0}-\eqref{eq:conformal_teukolsky_phi4}  and \eqref{eq:conformal_MasterFunc} show that the CEFE-Teukolsky equation can be written in terms of a single master function for ${}_\s\check{\phi}$ which reads:
\bea
&\big(\mathcal{A}_{TT}\partial_T^2 + \mathcal{A}_{T\varsigma}\partial_{T}\partial_\varsigma
+ \mathcal{A}_{T \varphi}\partial_{T}\partial_{\varphi}  + \mathcal{A}_{\varsigma \varsigma}\partial_\varsigma^2 
+ \mathcal{A}_{\varsigma\varphi}\partial_{\varsigma}\partial_{\varphi} \\
&  +\mathcal{A} \; \eth\bar{\eth} + \mathcal{B}_{\varsigma}\partial_{\varsigma} 
+ \mathcal{B}_{\varphi}\partial_\varphi + \mathcal{B}_T \partial_T + \mathcal{C}\big){}_{\s}\check{\phi}=0.
\eea
where
\bea
&&   \mathcal{A}_{TT} =  (2  - \dfrac{1}{4} \s^2) (2 F (1 + \kappa ^2 \varsigma ^2 (1  - T )^2)  - F^2 \kappa ^2 \sin^2\theta  (1  - T ) - \varsigma ^2 (-1 +  \nn  \\ && \varsigma  (1  - T )) (-1 + \kappa ^2 \varsigma  (1 - T )) (1  - T )) (1  - T ), \nn \\  
&&   \mathcal{A}_{T\varsigma}  =   2 (2  - \dfrac{1}{4} \s^2)  \varsigma F  (1 + \kappa ^2 (- F \sin^2\theta  + \varsigma ^2 (1  - T )) (1  - T )),  \nn \\
&&   \mathcal{A}_{T\varphi}  =  2 (2  - \dfrac{1}{4} \s^2) \kappa  \varsigma  (- F + \varsigma ^2 (1  - T )) (1  - T ),\nn \\ 
 &&  \mathcal{A}_{\varsigma\varsigma}  =  -F^2 (2 - \dfrac{1}{4} \s^2) \kappa ^2 \sin^2\theta  \varsigma ^2, \nn\\
 &&  \mathcal{A}_{\varsigma\varphi} =  -2 F (2 - \dfrac{1}{4} \s^2) \kappa  \varsigma ^2, \nn \\ 
&&   \mathcal{A} =  - (2-\dfrac{\s^2}{4})\varsigma^2\nn, \\ 
&&   \mathcal{B}_{\varsigma} =  F \kappa  \varsigma  (2i \s \cos\theta  \varsigma  + (-4  -4 s + \dfrac{1}{2} s^2) \kappa  \varsigma ^2 (1  - T ) \nn \\ && + \kappa  \sin^2\theta  (2 F \s  - (2  - \dfrac{1}{4} \s^2) \varsigma  F')), \nn \\ 
&& \mathcal{B}_{\varphi}  =   2 \kappa  \varsigma  (F \s + (-2  -2 s + \dfrac{1}{4} \s^2) \varsigma ^2 (1  - T )), \nn\\ 
&& \mathcal{B}_{T} = 
F^2 (4 + 2 s  - \dfrac{1}{2} \s^2) \kappa ^2 \sin^2\theta  (1  - T )  \nn \\ 
&&+ (2 + \s - \dfrac{1}{4} \s^2) \varsigma ^2 (2 + \varsigma  (-3 + \kappa ^2 (-3 +  4 \varsigma  (1  - T ))) (1  - T )) (1 - T ) \nn \\ 
&&  + F (-4  -2 \s + \dfrac{1}{2} \s^2 + 4 \kappa  \varsigma  (\dfrac{1}{2}i \s \cos\theta  + (-2  - \dfrac{3}{2} s   + \dfrac{1}{4} \s^2) \kappa  \varsigma  (1  - T )) (1  - T ) \nn \\ 
&&  + (-2 + \dfrac{1}{4} \s^2) \kappa ^2 \sin^2\theta  \varsigma  (1  - T ) F'), 
\nn\\
&&\mathcal{C}=  - F^2 \s (1 + \s) \kappa ^2 \sin^2\theta + 2 F \kappa  \varsigma  \bigg(-i \s^2 \cos\theta  + \s (1 + 2 \s) \kappa  \varsigma  (1  - T ) \nn \\ 
&&  + \dfrac{1}{2} \s \kappa  \sin^2\theta  F' \bigg)  + \varsigma ^2 (-2 \s + (3 \s + \dfrac{9}{4} \s^2) \varsigma  (1 + \kappa ^2 (1  -2 \varsigma  (1  - T ))) (1  - T )). \nn
\eea

Notice that the surface $\varsigma=0$ is a total characteristic as evidenced by the fact that
the Teukolsky operator evaluated there becomes the following transport equation for the master variable ${}_{\s}\check{\phi}$
\bea
\Big( (1-T)(1+T)\partial_{T}^2 - 2(1+\mathfrak{s})\partial_T  -\eth\bar{\eth}  - 2s \Big){}_{\s}\check{\phi} =0.
\eea
where $\s=2,0,-2$. Moreover, we also recover the equation for the massless scalar field derived in Ref.~\cite{Hennig:2020rns} when $\s=0$.

\section{Conclusions}\label{sec:Conclusion}
In this work we have derived a conformal counterpart of the Teukolsky equation within the framework of Friedrich’s conformal Einstein field equations (CEFEs), linearised around a Petrov-type D background. A key outcome of this work is the establishment of a bridge between the traditional curvature-based formulation of black-hole perturbation theory and the covariant conformal framework provided by the CEFEs.

\bigskip

As an explicit application, we specialised the CEFE-Teukolsky equation to the Kerr spacetime in hyperboloidal coordinates and recovered known results from the literature. In existing hyperboloidal derivations, the Teukolsky equation is typically obtained through effective coordinate-based constructions and ad-hoc rescalings for practical purposes, often deliberately avoiding a fully covariant conformal formulation in favour of accessibility and numerical tractability (see e.g.~\cite{Zenginoglu:2008uc,Zenginoglu:2011jz}). Here, we identify the hyperboloidal master variable geometrically as a Newman–Penrose component of the rescaled Weyl tensor. This makes explicit a covariant link between hyperboloidal approaches and the CEFE formalism.

\bigskip

We further applied the CEFE-Teukolsky equation to a conformal representation of the Kerr spacetime in which spatial infinity is realised as a blown-up cylinder. Although this construction differs from Friedrich’s original construcution of the cylinder at $i^0$ based on the $F$-gauge, it allows one to probe the neighbourhood of spatial infinity within a regular conformal setting, where the resulting equations reduce to transport equations along the cylinder. To the best of our knowledge, this constitutes the first formulation of a Teukolsky equation adapted to a cylinder representation of $i^0$.

\bigskip

The non-triviality of our result stems from the fact that the conformal factor is an independent variable in the full non-linear CEFE system. A central result is that the evolution equation for the perturbation of the rescaled Weyl tensor decouples from the perturbation of the conformal factor when linearising around Petrov-type D backgrounds. This decoupling relies on a highly nontrivial interplay between algebraic speciality and the conformal structure of the CEFEs: for generic backgrounds the perturbation of the conformal factor $\check{\Xi}$ is expected to couple to the remaining equations, and even within the Petrov-type D class additional couplings arise when considering variables beyond those captured by the conformal Teukolsky master functions or higher-order perturbations. Consequently, the decoupling obtained here should be understood as a distinctive feature of linear perturbations around algebraically special spacetimes.

\bigskip

The present results also clarify the relation between the hyperboloidal framework and the CEFE formulation. Within linear black-hole perturbation theory, hyperboloidal constructions typically adopt a prescribed conformal compactification adapted to the background spacetime, allowing the conformal factor to be treated as fixed at the level of the effective equations. The CEFE formulation, by contrast, promotes the conformal factor to an independent variable. The absence of conformal-factor coupling at linear order for Petrov-type D backgrounds therefore provides a geometric explanation for the agreement between hyperboloidal formulations and the CEFE-Teukolsky equation in this regime, while also indicating how additional couplings naturally emerge once one moves beyond the leading-order curvature dynamics. In this sense, the CEFE framework may be viewed as providing a geometric completion of the conformal structures implicitly exploited in hyperboloidal perturbation theory, embedding effective constructions within a fully covariant and regular formulation.

\bigskip

Establishing a conformal formulation of the Teukolsky equation consistent with traditional approaches represents only a first step toward a complete, self-consistent, conformal and geometrically regular treatment of black-hole perturbations. Looking forward, an important direction concerns metric reconstruction, potentially through variables intrinsic to Friedrich’s formulation of the CEFEs. While perturbations are naturally expressed at the level of curvature quantities, gauge issues inevitably arise in reconstructing the metric. Understanding how existing reconstruction strategies (e.g. \cite{Green:2019nam, Hollands:2024iqp} and references therein), as well as Kerr–Schild-based approaches for addressing nonlinearities \cite{Harte:2014ooa,Harte:2016vwo}, integrate with the CEFE formalism remains an open problem. In this context, recent developments in conformal formulations of the Geroch–Held–Penrose (GHP) formalism~\cite{Schneider:2025oqq} suggest a promising avenue for reformulating perturbation theory in a conformally covariant tetrad language, potentially offering a natural interface between CEFE-based approaches and the standard curvature formulations used in black-hole perturbation theory, as well as a pathway toward extensions beyond the first order and linear regime.

\section*{Acknowledgments}
We thank Anıl Zenginoğlu and Juan A. Valiente Kroon for helpful discussions on the topic and comments on this research program. RPM and JCF thank the Yukawa Institute for Theoretical Physics (YITP), Kyoto University, which hosted the Gravity and Cosmology 2024 long term workshop where foundational discussions for this work took place.
RPM acknowledges support by VILLUM Foundation (grant no. VIL37766) and the DNRF Chair program (grant no. DNRF162) by the Danish National Research Foundation.
The Center of Gravity is a Center of Excellence funded by the Danish National Research Foundation under grant No. 184.
This project received financial support provided under the European Union’s H2020 ERC Advanced Grant “Black holes: gravitational engines of discovery” grant agreement no. Gravitas–101052587, as well under the research and innovation programme  Marie Sklodowska-Curie grant agreement No 101007855 and No 101131233.
Views and opinions expressed are however those of the author only and do not necessarily reflect those of the European Union or the European Research Council. Neither the European Union nor the granting authority can be held responsible for them. JCF acknowledges support by the European Union and Czech Ministry of Education, Youth and Sports through the FORTE project No. CZ.02.01.01/00/22\_008/0004632.

\section*{Bibliography}
\bibliographystyle{unsrt.bst}

\end{document}